**Master's Thesis**

# Towards an Autonomous System Monitor for Mitigating Correlation Attacks in the Tor Network

Supervisor: Professor Yasuhito Asano
Department of Social Informatics
Graduate School of Informatics
Kyoto University

Nguyen Phong HOANG

August 1st, 2016

# Towards an Autonomous System Monitor for Mitigating Correlation Attacks in the Tor Network

Nguyen Phong HOANG

**Abstract**

As we are living in the digital age, information technology appears in nearly every aspect of our daily lives, and plays an integral role in agriculture, industry, education, business, politics, etc. Thanks to its undeniable convenience, information technology has helped to make our lives easier and more comfortable. For instance, VoIP (voice over IP) provides a cheaper and smoother experience for intercontinental voice communications. Moreover, Internet advancement continues to provide easier ways to store, share, retrieve, and access an infinite amount of information online. However, despite the many ways that information technology helps to enrich our lives, it functions as a double-edged sword, with the potential to become a trap for users with a limited understanding of technological complexities.

Along with the rapid expansion of the Internet and various forms of information technology in recent years, the issues of censorship and surveillance in cyberspace have become more prevalent. Many governments around the world are making use of information technology to extensively monitor and censor the Internet. Even in areas with a long history of freedom of speech, like North America and Western European countries, surveillance of Internet user activity occurs on a daily basis. Since most Internet users are unaware of the extent to which they are being monitored online, they are susceptible to having their privacy compromised by such unconstitutional surveillance activities. Through this study, the author would like make Internet users more aware of the issues surrounding online privacy. In addition, several privacy-enhancing technologies will be introduced, along with their advantages and disadvantages, to educate Internet users about the tools that can be used to help protect their privacy in cyberspace.

Among the available pro-privacy tools, Tor (The Onion Router) is the most popular and will be the focus of this study. By carefully investigating its operating mechanism and remaining weaknesses, the author shows that Tor is not as safe as many Internet users believe. Tor cannot guarantee complete anonymity while surfing the Internet. By using a real-life example, the author shows that there are cases in which Tor is ineffective for this purpose. Tor remains vulnerable to a type of attack known as a correlation attack, which can be carried out by entities that have the ability to observe a large portion of Internet traffic, especially the autonomous system (hereafter: AS).

Regardless of the many AS-aware Tor clients proposed in studies over the last decade to help users mitigate the threat of AS-level adversaries, none of these clients have been widely adopted by end users.

After carefully considering the scalability problem in Tor and exhaustively evaluating related works on AS-level adversaries, the author proposes ASmoniTor, which is an autonomous system monitor for mitigating correlation attacks in the Tor network. In contrast to prior works, which often released offline packets, including the source code of a modified Tor client and a snapshot of the Internet topology, ASmoniTor is an online system that assists end users with mitigating the threat of AS-level adversaries in a near real-time fashion. For Tor clients proposed in previous works, users need to compile the source code on their machine and continually update the snapshot of the Internet topology in order to obtain accurate AS-path inferences. On the contrary, ASmoniTor is an online platform that can be utilized easily by not only technical users, but also by users without a technical background, because they only need to access it via Tor and input two parameters to execute an AS-aware path selection algorithm.

With ASmoniTor, the author makes three key technical contributions to the research against AS-level adversaries in the Tor network. First, ASmoniTor does not require the users to initiate complicated source code compilations. Second, it helps to reduce errors in AS-path inferences by letting users input a set of suspected ASes obtained directly from their own traceroute measurements. Third, the Internet topology database at the back-end of ASmoniTor is periodically updated to assure near real-time AS-path inferences between Tor exit nodes and the most likely visited websites. Finally, in addition to its convenience, ASmoniTor gives users full control over the information they want to input, thus preserving their privacy.

**Towards an Autonomous System Monitor for Mitigating Correlation Attacks in The Tor Network**

# Contents

# Chapter 1 Introduction

First of all, this chapter aims to provide an overview of privacy, Internet privacy and privacy-enhancing technologies. As a major focus of this study, The Onion Router (Tor) is then introduced and discussed comprehensively, including its design, operating mechanism, and drawbacks. Finally, the primary motivation of this study and the overall structure of the thesis are described at the end of the chapter.

## 1.1 Privacy

### 1.1.1 Background

Humans have been developing strategies to protect confidential information for many years, even prior to the digital age. The use of codes and ciphers is a common example [1]. Confidential information, also known as private or secret information, can belong to a single person or a group of people. This information can be anything, ranging from biological information or healthcare information, to financial transactions or geographic location data.

However, due to differences among cultures and individuals, there is no concrete definition of confidentiality, privacy or secrecy. For instance, biographical, location and gender data for closeted GLBT people is sensitive information. On the other hand, this information may serve as valuable tool for GLBT activists trying to raise public awareness of the GLBT community. As a result, the idea of adequate privacy preservation is something that is highly variable from person to person [2].

Despite the extensive variability in definitions of privacy, there is some commonality that exists: sensitivity and importance of the information to its owner. Because the leak of such information may have a negative impact on the owner, the right to privacy is rigorously protected by all of the Western European countries in Article 8 of the European Convention on Human Rights [3], and is clearly declared by the United Nations in Article 12 of The Universal Declaration of Human Rights [4].

**1.1.2 Privacy on the Internet**

Since the invention of electronic computers, humans have witnessed the rapid development of the Internet, which is a continuously expanding network of interconnected computers and other electronic devices around the globe. The evolution of the Internet has significantly changed the way by which people exchange information and communicate with one other. In that context, the ability of an Internet user to preserve her privacy depends on what information she wants to reveal or conceal while using the Internet. However, privacy on the Internet, although globally discussed and articulated in law, is not completely preserved. The nature of the Internet and the advancement of technology have made it easy to collect private information [5]. In most cases, Internet users (especially non-technical users) do not know how to control and are unaware of the potential for extensive collection of personal information by third parties without the user's consent.

Privacy on the Internet can either encompass personally identifiable information (such as biographical information, an IP address of a connecting device, and so on) or non-personally identifiable information (such as mouse clicks, or web surfing behaviors). The combination of such information, although not complete, is often unique enough to identify a specific person. For clarity, the following terms are often introduced when discussing privacy on the Internet: untraceable, unobservable, anonymous.

- Untraceable is when a piece of information cannot be used to trace other information of the Internet user. For instance, an attacker can track down the location of a victim using an IP address alone. Therefore, by hiding the real IP address, the Internet user can reduce this risk to a certain extent, depending on which pro-privacy technology is used.

- Unobservable is when the Internet user cannot be observed. For example, Internet service providers (hereafter: ISP) cannot monitor the content of the data packets sent over their network if the Internet user encrypts the connection by means of cryptography.

- Anonymous is when the Internet user does not leave any clue of personally identifiable information while online.

### 1.1.3 Privacy-enhancing technologies

Due to the increasing concern about privacy in the digital age, many privacy-enhancing technologies (hereafter: PET) have been invented to support the Internet user. PET is a suite of computer applications and algorithms that, when combined with other online services or applications, protects the privacy of Internet users. In other words, PET can be defined as "a system of ICT measures protecting informational privacy by eliminating or minimising personal data thereby preventing unnecessary or unwanted processing of personal data, without the loss of the functionality of the information system. [6]"

Although there are many technologies that can be employed to effectively protect one's privacy, the scope of this study is limited to online privacy and anonymity, thus only those techniques that are widely used by Internet users are discussed. These techniques include the private browsing function integrated into most Internet browsers, proxies, virtual private networks (VPN), and The Onion Router (Tor).

- Private browsing mode of Internet browsers

Most Internet browsers have implemented a built-in private browsing mode to protect user privacy. This mode is also known as "Incognito mode" in Chrome, "Private Browsing mode" in Firefox, and "InPrivate mode" in Microsoft Edge. The advantage of this browsing mode is that it deletes all cookies and browsing history when the user closes the browser. However, like normal browsing mode, it does not help to protect the Internet user from being monitored by man-in-the-middle attacks, or from the collection of their private information by accessing the destination server.

- Proxy servers

A proxy server prevents the destination server from logging the real IP address and other related information about the device being used by the Internet user. However, transmitted data packets still have to go through the physical network of the ISP before being sent to the proxy server, thus it is still vulnerable to the man-in-the-middle attack. In other words, anyone who can monitor the traffic between the user and the proxy can still capture the transmitted data packets and analyze them to uncover private information about the user.

- Virtual private network (VPN)

VPN was introduced to solve the key problem of proxy servers mentioned above. Unlike proxy servers, it encrypts all of the packets sent from the user to the VPN server. The VPN server, on behalf of the user, connects to the intended destination. The advantage of VPN is that once it is enabled, all traffic will be encrypted, regardless of the type of application being used. As a result, the ISP or the man-in-the-middle cannot uncover private information from captured data packets. Nonetheless, VPN cannot guarantee the Internet user's privacy. There have been several cases in which VPN providers have made businesses out of users' private information, or have made the logged data available to law enforcement upon request.

- The Onion Router

Tor was created to mitigate the drawbacks of the aforementioned pro-privacy technologies. It employs asymmetric cryptography and creates multiple layers of encryption. When transmitting packets between the user and a final destination host, a random path consisting of three nodes is used so that each node only knows a part of the whole transmission process, guaranteeing the anonymity of the user. Tor has been proved to be the most robust pro-privacy tool in previous work [5]. Since it is the dominant topic throughout this study, details of its design and operating mechanism will be discussed comprehensively in Section 1.2.

Table 1 summarizes the pros and cons of each technology, as studied in previous work [5].

Table 1: A comparison of pro-privacy technologies

| Testing tool | Private Browsing mode | Proxy | VPN | Tor |
|---|---|---|---|---|
| Against IP logging | Fail | Pass | Pass | Pass |
| Against man-in-the-middle attack | Fail | Fail | Pass | Pass |
| Against trace-back | Fail | Fail | Fail | Pass |
| Support of dynamic IP and change of transmission path | No | No | Limited | Yes |
| Cost | Free | Flexible | Flexible | Free |
| Pro-privacy level | Low | | | High |

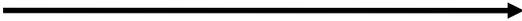

## 1.2 Tor: The Onion Router

### 1.2.1 Design goals

Nowadays, it is widely known that government surveillance activities are happening a daily basis [7], [8], even in regions with a long history of freedom of speech like Western countries. According to the Guardian, the NSA has a program called Marina that can track and store the online metadata of millions of Internet users for up to one year [9]. It does not only take place in the United States, but also in other countries like Japan [10]. At the same time, Britain's GCHQ has secretly gained access to the networks of cables that transmit the world's phone calls and Internet traffic [11].

Because of this widespread surveillance activity, anonymous communication has drawn remarkable attention from both researchers and ordinary Internet users, and it grows to support an increasing number of users, with various applications, who want to protect their privacy while online [12].

As discussed in subsection 1.1.3, despite the existence of many pro-privacy and anti-censorship applications that are freely available on the Internet to protect the Internet user's privacy and anonymity (proxy servers, VPN, etc.), Tor has proved to be the most robust tool among them. It helps Internet users prevent those entities with the ability to monitor Internet connections from discovering what sites are being visited, and it also prevents the visited sites from knowing the real IP address of Internet users. For instance, people in China may use Tor (with the help of Tor's Bridge nodes, since the direct access to Tor is also blocked in China) to access websites that are censored by the Great Firewall of China, such as BBC news, Facebook and YouTube, while journalists and activists use Tor to protect themselves from the surveillance of repressive regimes.

### 1.2.2 Operating mechanism

The Tor network has two main elements: a group of nine directory authorities and a network of approximately 7000 relays [13]. The directory authorities are centralized, trusted servers that keep track of the whole Tor network. The relays (aka routers or nodes) are voluntary computers, distributed around the world, and can be categorized into bridge, guard, middle and exit relays based on their functionality and position in the virtual circuit.

As shown in Figure 1, the Tor user software (a Tor client which can be a Tor Browser Bundle [14], Tails [15], or any interface that talks to the Tor network) initially fetches a consensus "network status" document containing all Tor relays' information from hard-coded directory authorities. Before being sent to the intended destination host, data packets are encrypted multiple times using public-key cryptography. These packets are then forwarded through a virtual circuit, made up of several relays, using Onion Routing Protocol [16]–[19]. (Ideally three relays for direct users and more in the case of bridge users [20]).

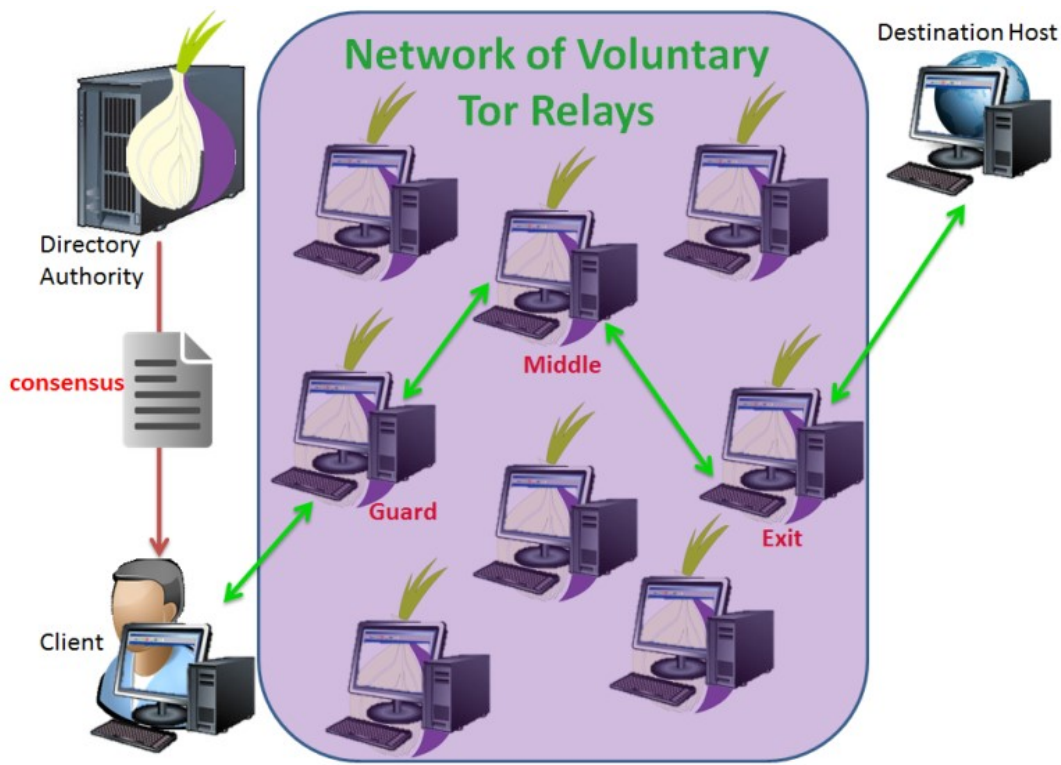

Figure 1: System Architecture of the Tor network

As illustrated in Figure 2, at each relay, one encryption layer of the packet is decrypted to reveal the next relay's IP address until the packet reaches the final destination. Later, response packets are sent in a reverse fashion back to the client. The whole mechanism is similar to peeling away the layers of an onion, where each relay only knows the previous and the next relays. By this mechanism, without traffic analysis, none of the relays in the virtual circuit can correlate the original sender and the destination. Thus, the privacy of the client is guaranteed.

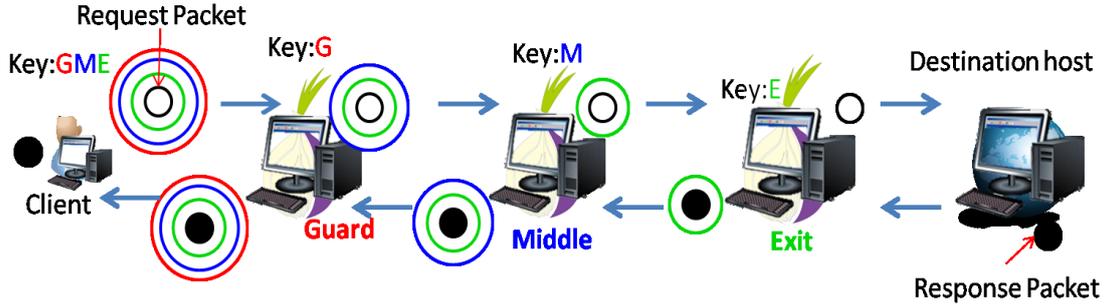

Figure 2: Packet transmission in the Tor network

In recent versions, in order to improve both performance and overall traffic balance in the Tor network, each client selects relays in proportion to their bandwidth to form the virtual circuit. For instance, given a total number $N$ of routers in the network and the bandwidth $b_i$ provided by router b, the probability that router b will be chosen is calculated as follows:

$$\frac{b_i}{\sum_{k=1}^{N} b_k}$$

In addition, for security purposes, Tor client does not choose the same router twice for the same transmission circuit. Similarly, relays in the same /16 subnet (aka Class B network) are not selected to be in the same circuit. For instance, Tor client will choose only one relay from the IP range XXX.XXX.0.0 to XXX.XXX.255.255 if there is more than one relay in the IP range. Finally, relays in the same family are not selected to be in the same circuit. Two relays are in the same family if each one lists the other in the family entries of its descriptor [21].

### 1.2.3 Weaknesses

Despite its popularity, Tor has some drawbacks that have been identified. These drawbacks can be clustered into two groups: scalability and security. In contrast to previous works that addressed these problems in a theoretical manner, this study investigates the problem of scalability based on real-life data so that readers can gain a more comprehensive understanding of the remaining scalability problem in current Tor design. In addition, a set of well-known and severe attacks against Tor will be revisited briefly in this section.

**Scalability**

In order for Tor to function properly, all Tor clients try to continually maintain a live consensus network status document containing information about all relays that is necessary for the client to build anonymous circuits. As defined in Tor directory protocol, a network status document is "live" if the time in its "valid-until" field has not passed [22]. At the time of writing this thesis, Tor directory authorities publish a new consensus network status document every hour, and that consensus is valid for three hours after its release. As a result, all Tor clients try to update to a new consensus network status document every few hours (less than 3 hours) to ensure their information about all available relays is current.

Nevertheless, McLachlan *et al.* [23] had foreseen that the Tor network would be spending more bandwidth on serving the consensus network status document because of the dramatic increase in the number of both relays and clients. In more detail, given $n$ clients and $r$ relays in the Tor network, the bandwidth cost for distributing the consensus network status document to all clients in recent Tor design is estimated by $O(nr)$, while the bandwidth available for the whole network only increases by $O(r)$.

For a more comprehensive insight into how the number of clients and relays impact the bandwidth overhead for serving the consensus network status document, past data of direct users and relays were retrieved from Tor Metrics [24] and plotted in Figure 3 and Figure 4.

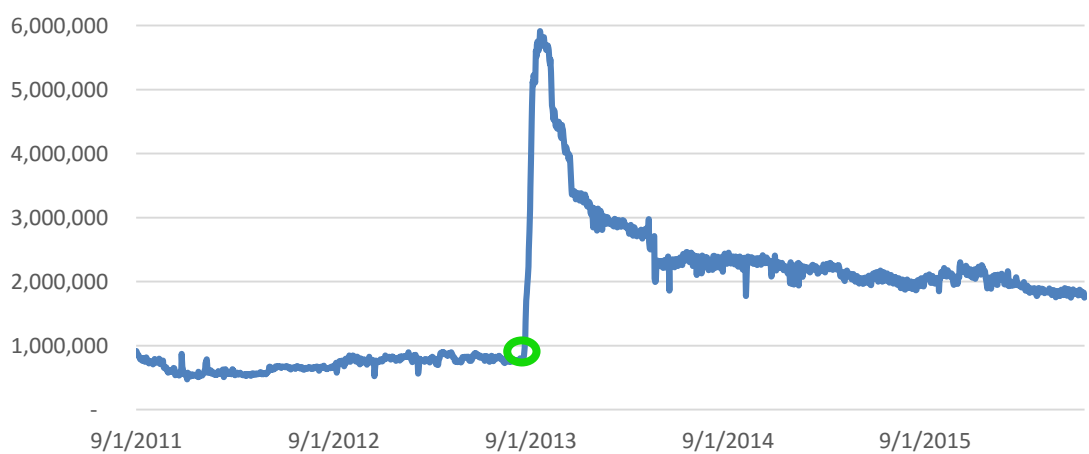

Figure 3: Daily direct users (in million)

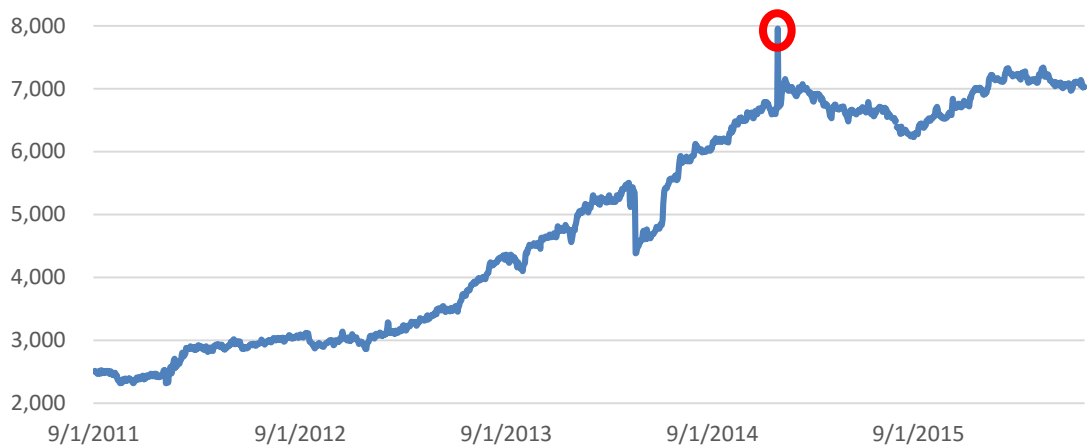

Figure 4: Relays in the Tor network over time

As shown in Figures 3 and 4, the number of users has doubled, while the number of relays has nearly tripled in the last 5 years. As highlighted by the green node in Figure 3, starting around mid-August 2013, Tor was under a coordinate attack by a bot twice the size of the regular Tor network. This attack is the reason there were about 6 million daily users at that moment [25]. Tor was still usable at that time but could not avoid wasting a huge amount of bandwidth (approximately 18 TB/day was the peak of the attack, as highlighted by the green node in Figure 5) on answering the directory requests (aka request to download the consensus network status document), and some relays were saturated by the huge number of connections and circuits that they had to handle.

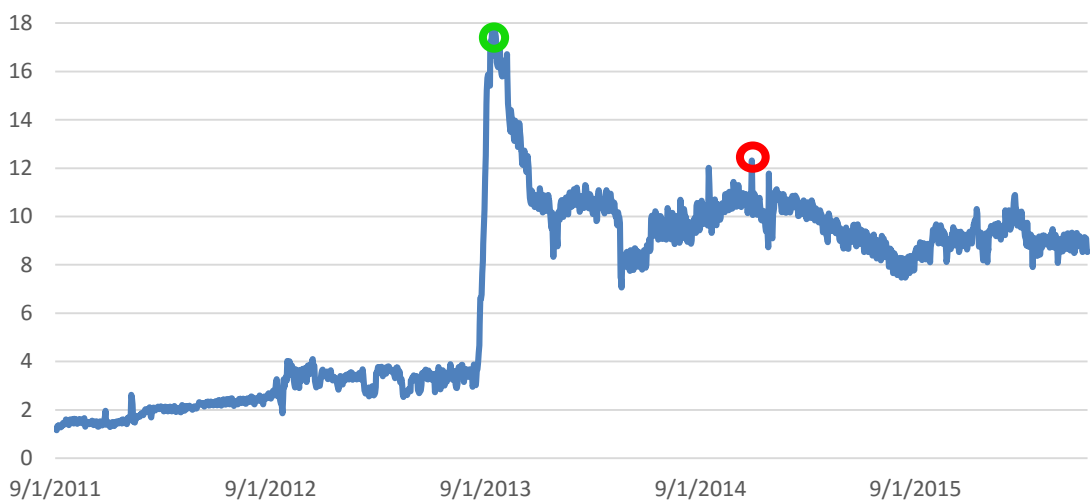

Figure 5: Bandwidth spent on serving directory requests (in TB/day)

In addition, during December 2014, there was a Sybil attack [26] on Tor in which a large number of small exit relays were created. It was detected immediately [27], but still caused an increase in the size of the consensus network status document, as highlighted by the red nodes in Figures 4, 5 and 6.

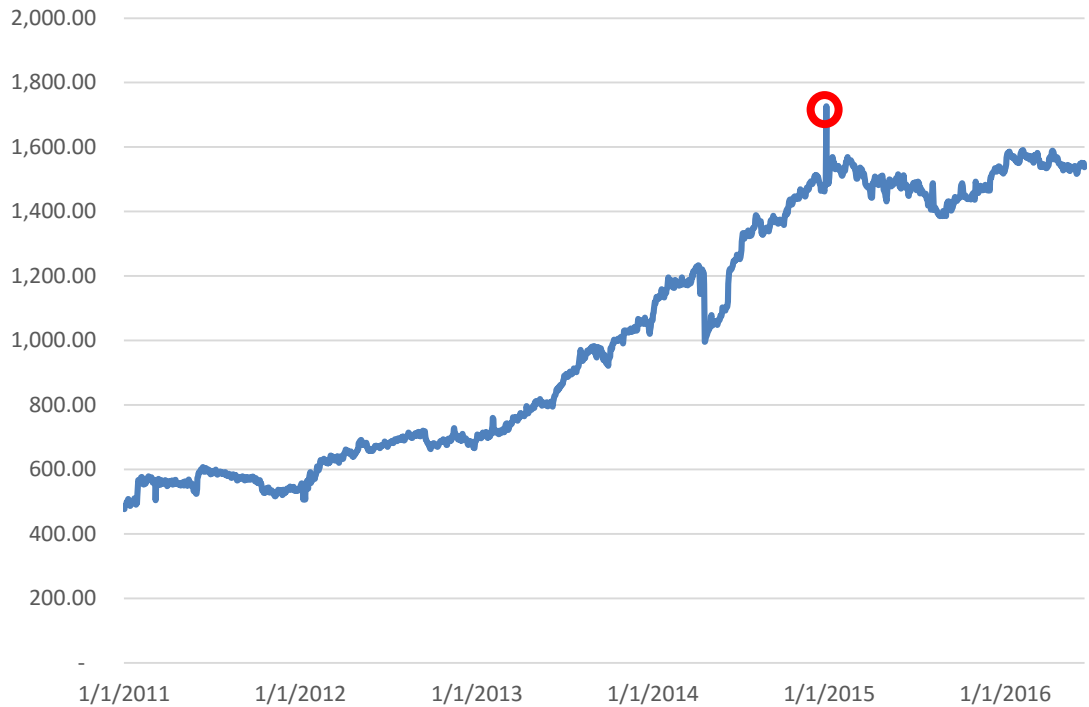

Figure 6: Size of the consensus network status document (in KB)

Figures 5 and 6 show the sensitivity of the cost of bandwidth for serving the consensus network status, particularly to changes in the number of clients and relays in the Tor network. The data for Figure 5 and Figure 6 are not readily available; therefore, we first fetched the raw data of *number of bytes spent on answering directory requests* from Tor Metrics [28] and *past data of consensus files* from Tor Collector [29]. Then, some calculations and parsing procedures were conducted to retrieve necessary values for plotting the graphs. The size of consensus file in Figure 6 is the average size of consensus files collected per day.

By observing Figures 3 through 6, there are some obvious correlations in the increasing pattern between Figures 3 and 5, and between Figures 4 and 6, while Figures 4 and 5 have a similar pattern during the period from March 2014 until June 2016. As a result, we can preliminarily conclude that both number of clients and relays have significant effect on bandwidth cost for serving the consensus network status document. However, the number of clients has a more obvious effect on the bandwidth cost compared to other factors. To that end, there are four possible combinations of changes that could happen to the number of clients and relays in the future. The likely impact of these changes on bandwidth cost is shown in Table 2.

Table 2: Possible changing patterns of bandwidth cost for serving the consensus network-document in the Tor network

|   | Number of clients | Number of relays | Bandwidth cost for serving the consensus network-document |
|---|---|---|---|
| 1 | decrease | decrease | will decrease |
| 2 | decrease | increase | decrease will more likely happen than increase |
| 3 | increase | decrease | increase will more likely happen than decrease |
| 4 | increase | increase | will dramatically increase |

The number of Tor clients is predicted to continue increasing in the coming years due to deep concerns about censorship and surveillance activities in cyberspace [30]. As a result, the two highlighted patterns in Table 2 are more likely to happen. Hence, future works need to take the scalability into careful consideration and restrain from unnecessarily increasing the size of the consensus network status document or requiring clients to download more consensus documents. Because most of Tor's relays are voluntarily operated [31], or run by limited sources of donation [32], bandwidth must be saved and used in an economical fashion. Therefore, all of the proposals in this study will be discussed and suggested in a careful manner, such that they will not cause additional scalability problems to the entire Tor network.

In addition to the above findings, dividing data in figure 6 by the data in figure 4 and taking the average value gives a better representation of the size of the consensus document. With recent Tor design, the result shows that every new relay joining Tor would increase the size of the consensus network status document by approximately 210~230 bytes, regardless of its bandwidth resource. Although discussing the contribution of fast/slow relays to the overall Tor network is beyond the scope of this study, it is vital to argue that relays with higher bandwidth would be preferable to the slow ones. Because of the recent Tor's relay selection algorithm, very little traffic will be directed to a slow relay, while the whole network still needs to maintain and distribute information for that particular relay (e.g., bandwidth cost for periodically scanning the relay's bandwidth, computational cost for voting the relay among directory authorities and bandwidth cost for distributing the relay's information to the entire Tor network). It should also be noted that the importance of those slow relays is undeniable in the Tor network because anonymity is the direct result of diversity in the number of relays, users and so on. Therefore, to improve the aforementioned circumstance, Tor developers suggest that a relay should have a minimum bandwidth of 250 KB for both download and upload links [33].

**Low Latency Attack**

This type of attack is also known as a traffic analysis attack, or a congestion attack. With only a partial observation of the Tor network, Murdoch and Danezis in [34] were able to make use of traffic analysis techniques to predict which Tor nodes were being used to relay an anonymous stream, thus allowing them to compromise the identity of the client. The attack is based on the fact that Tor maintains a bin for each anonymous stream and relays them in a round-robin fashion. As a result, the latency (i.e., the gap of time between bin accesses) is a function of the number of bins. Thus, adding an anonymous stream to a given Tor node affects the latency of all other stream being transmitted via that node. The adversary in this attack can be a bad destination host that wants to reveal the identity of its clients. When a client connects to that host via Tor, the host tries to respond the client's request with manipulated packets that makes the traffic have a very specific pattern. By this method, the adversary can learn the nodes on the Tor path by probing the suspected routers using another Tor client. The attack was first reported in 2005, when there were only 50 nodes in the Tor network. However, the increasing number of Tor clients and relays in its network has made the original congestion attack infeasible, thus it is no longer employed to attack in today's Tor network. In 2009, Evans *et al.* in [35] proposed an enhanced version of the original low-latency attack by injecting JavaScript code into the client's request. As a consequence, those clients whose Tor browser has JavaScript still enabled will be detected by the specific traffic pattern manipulated by malicious JavaScript. The main purpose of the attack is to identify the guard node of the client, thus compromising her identity.

**Path Selection Attack**

As mentioned in section 1.2.2, in order to provide a low-latency service, and also to improve the overall traffic balance in the Tor network, Tor path selection algorithm is done in proportion to the bandwidth of the relay to form the virtual circuit (aka anonymous path).

However, Øverlier and Syverson found that original Tor path selection algorithm was vulnerable to manipulation [36]. Since the original Tor lets the relays self-report their bandwidth, the adversary can create a set of low-resource relays and maliciously advertises high-bandwidth capacities, thus increasing the probability that his malicious relays will be selected by Tor client. In a worse case, if many relays under the control of the adversary are selected, a circuit-linking algorithm or a timing attack can be deployed to compromise the identity of the client. In [37], the authors show that with the control of only 6 out of 66 relays run in a PlanetLab-based [38] Tor experiment, they could compromise 46% of anonymous paths.

To remedy the problem, Tor has been deploying bandwidth authorities to measure the bandwidth of all of the relays in the Tor network since 2009.

**Passive Attack**

Fingerprinting attacks and AS-level attacks are two types of passive attacks.

A fingerprinting attack [39]–[41] is a type of attack in which the adversary only needs to monitor the traffic between the client and the guard relay. The success of this attack is heavily based on machine learning techniques (e.g., classifying, clustering, etc.) To carry out the attack, the adversary first studies the traffic pattern of a set of websites of his interest, using supervised machine learning techniques in order to create a classifier. The adversary then observes the traffic between the client and the guard node, letting the classifier predict whether or not the Tor client is browsing one of the websites of the attacker's interest. Some previous works suggested adding noise to the traffic to alleviate this type of attack [42]–[47].

An AS-level attack is also known as a correlation attack. This type of attack is when an adversary can monitor both ends of a Tor path. In other words, the adversary needs to observe the traffic between the client and the guard relay, as well as the traffic between the exit relay and the intended destination host. By analyzing the time, size and frequency of packets passing through the two ends, the adversary will be able to correlate the client and her accessing destination host, thus deanonymizing the Tor user. Although many studies and proposals [48]–[55] have been done to weaken this attack, it still remains a lingering problem of Tor due to the nature of the Internet including centrality, routing policy, and the general lack of awareness among non-technical Tor users. Since the scope of this thesis mainly focuses on how to efficiently and effectively mitigate the threat of AS-level adversaries, Chapter 2 is dedicated to a comprehensive discussion of previous related works in the domain of AS-level attacks against Tor.

## 1.3 Contributions of this study

Three specific contributions of this study include the following:

- Based on historical data of the Tor Network, we are able to measure the increasing size of the consensus file caused by a new relay (if added) to the Tor network, which varies from 210 KB to 230 KB.
- We examine the feasibility of the Shorter-Tor-path approach to neutralize the risk of the AS-level attack.
- Finally, we are working toward realizing a system called ASmoniTor to assist the Tor user in mitigating the threat of being deanonymized by AS-level adversaries.

## 1.4 Thesis organization

The thesis is organized in 6 chapters. The First two chapters are dedicated to general introductions to online privacy, The Onion Router and related works. Chapter 3 revisits the scalability problem in Tor and preliminarily introduces a proposal to mitigate AS-level adversaries. Chapter 4 examines the feasibility of the shorter-Tor-path approach. In Chapter 5, we introduce ASmoniTor, which is an online system to assist Tor users in combatting AS-level correlation attacks. The conclusion of this thesis and future works are summarized in Chapter 6.

# Chapter 2 Related Work

Tor is different from high-latency anonymity techniques, such as Mixminion [56], which add some delay into packets to combat very powerful adversaries who are able to monitor a large portion of Internet traffic. In contrast, Tor provides a low-latency anonymity service for real-time interactive applications, such as web surfing. As an inevitable consequence, adversaries with the ability to monitor traffic at the two ends of a Tor communication (i.e., the route between the client and the guard relay, and the route between the exit relay and the final destination) can deploy traffic analysis techniques to correlate the size and transmitted time of data packets in order to deanonymize the user. This chapter is dedicated to a discussion of related studies on this lingering problem.

## *2.1* Nick Feamster and Roger Dingledine [48]

As one of the first studies conducted to empirically analyze the threat of AS-level adversaries in the Tor network, this study proposed a metric called *Location Independence* to measure the probability that the incoming and outgoing connections (i.e., the connection between the client and the guard node, and the connection between the exit node and the intended destination host) could have passed a common AS. The authors found that a single AS could observe both connections 10%~30% of the time.

However, at the time of conducting the study, there were only 33 relays in the Tor network. In addition, the authors carried out the experiment based on personal opinions about the websites that Tor users often visited, and only took into account the Tor clients located in the United States.

## 2.2 Matthew Edman and Paul Syverson [49]

Five years after the work of Feamster and Dingledine [48], at a time that when the Tor network had grown to have around 2000 relays and 250000 users, the problem of AS-level adversaries was revisited. In spite of the growth, the authors in [49] showed that the threat of a single AS monitoring traffic between the two ends of anonymous Tor paths was greater than it was in previous work, with the ability to observe 39.4% of all randomly generated anonymous paths.

However, one of the drawbacks of the study is that the authors measured the AS threat based heavily on an AS path inference technique [57], which often generates paths different from the real-time traceroute result and does not accurately reflect the threat of AS-level adversaries. The difference between paths collected by real-time traceroute measurements and by inference is alleged to be 80% in a current study [53], thus necessitating a more reliable method.

Furthermore, in order to obtain samples of the clients' ASes and the likely visited websites' ASes, the author created two Tor nodes (one guard and one exit) to collect traffic information. However, this method is no longer reliable to measure the risk of AS-level adversaries in the current Tor network because the network has expanded to more than 2000 guard nodes and 800 exit nodes, as we will discuss in subsection 4.2.2. In addition, the issue of ethics has to be carefully considered when collecting statistical data in the live Tor network [58]. In other words, measuring sensitive data can be harmful to the anonymity of Tor users, especially when researchers publish the collected data. Even when the data is not published in an obvious fashion, the adversary may still be able to carry out some linkage attacks, based on their background knowledge of the users. To that end, doing Tor experiments directly on the live Tor network is not an appropriate direction in Tor research.

## 2.3 LASTor [50]

Akhoondi *et al.* designed a Tor client called LASTor, which is claimed to be able to significantly reduce the latency of data transmission by 25%, compared to the default Tor client, by choosing the anonymous Tor path based on the Tor nodes' geolocation information. In addition, the authors developed an algorithm, which is said to be able to detect paths on which an AS can correlate traffic between two ends. At the end of their paper, the authors announced their plan to publish the LASTor client for public use.

However, at the time of conducting this study, the LASTor client has not yet been made available to the end users. Implementation of the proposal was likely unfeasible because the authors did not carefully consider the issue of traffic bottlenecking of the Tor network in their proposal.

## 2.4 Aaron Jonson *et al.*[51]

Among many specific contributions of this study, the creation of TorPS, a realistic Tor path simulator, allowed Jonson *et al.* to calculate the time needed for an AS or IX (Internet Exchange point) to deanonymize an anonymous circuit. The authors took into account 3 ASes and IX adversaries, as observed in their data, and found that the chance of an AS compromising the identity of the Tor user by a correlation attack is 1.6% for the top 3 ASes in their study.

## 2.5 RAPTOR[52]

Although the threat of AS-level adversaries has been discussed for more than a decade, Sun *et al.* recently introduced RAPTOR, which is a suite of renewed attack techniques that can be utilized by an AS-level adversary to deanonymize Tor users. Since Tor communication is done with Transmission Control Protocol (TCP), the authors found that an AS-level adversary on the reverse path of a transmission can make use of TCP acknowledgment packets to strengthen its correlation attack on the privacy of Tor users.

In more detail, RAPTOR encompasses three small attack techniques:

1. Exploiting the asymmetric nature of the Internet to conduct traffic analysis and correlation attacks at both ends of the Tor communication.
2. Exploiting the dynamic change in the Internet topology to increase the chance that a particular adversarial AS can appear simultaneously on both ends of the Tor communication.
3. Manipulating Internet routing policy through BGP hijacking (to locate guard relays used by the client) and BGP interception (to perform traffic analysis).

With these techniques, the group was able to deanonymize Tor users with up to 95% accuracy by using asymmetric traffic correlation exclusively.

Although the group was successful in demonstrating the RAPTOR attacks, the method they applied to model Tor users and intended destination hosts is not reliable because they only used 100 nodes hosted in the PlanetLab [38] network, with 50 nodes playing the role of Tor clients, and the other 50 nodes playing the role of intended destination hosts. In our opinion, this method is not sufficient to represent real clients and the most likely visited websites in the current Tor network. As a result, although the RAPTOR attack techniques may work to some extent when employed in the live Tor network, the statistical results of this study are questionable.

## 2.6 Joshua Juen *et al.* [53]

The work by Juen *et al.*, although not new, again showed that AS-level routing predictions should not be heavily relied upon for creating AS-aware Tor clients to mitigate AS-level adversaries. The authors found that AS paths predicted by AS inference techniques differ from traceroute measurements 80% of the time.

Nonetheless, one of the drawbacks of this study is that the authors passively relied on traceroute results submitted by a limited group of only 28 volunteers who operate Tor exit relays, while there are more than 800 exit nodes in the Tor network, as we will mention in subsection 4.2.2. Furthermore, the authors did not reasonably investigate the websites that are likely visited by Tor users. Instead, they randomly ran the traceroute measurements from 28 volunteers' Tor exits to an IP picked up within each of the approximately 500,000 prefixes acquired from the September 2013 *Routing Information Bases* of the *Route Views* routers. Thus, their findings may not be an accurate representation of the true threat posed by AS-level adversaries.

## 2.7 Astoria [54]

Nithyanand *et al.* performed an empirical study to measure the threat of AS-level adversaries. They found that up to 40% of Tor paths generated by the current Tor path selection algorithm are susceptible to AS-level adversaries, 42% in the case of colluding adversarial ASes, and 82% in the case of state-level adversaries.

The group then introduced Astoria, an AS-aware client, which is alleged to be able to reduce the above threats to 2%, under 5%, and 25% for AS-level adversaries, colluding adversaries ASes, and state-level adversaries, respectively.

Although the source code Astoria was published last year, it has not been widely adopted. Perhaps, it is because the process of compiling source code of a large packet (about 462 MB after extracted) is too complicated for end users, particularly non-technical and non-Linux users.

## 2.8 Cipollino [55]

The study conducted a historical analysis to measure how the threat from AS-level adversaries has changed over the past five years. Similar to [49], the authors showed that the threat posed by AS-level adversaries has grown, despite the dramatic increase in number of Tor relays over the last decade. For the use of interactive web surfing, the authors found that 31% of the anonymous paths are vulnerable to AS-level adversaries. For the mixed use of many applications simultaneously, such as Web, Bit Torrent, IRC, and email, 30% of the paths were found to be vulnerable. In addition, the authors proposed Cipollino, which is an improved version of Astoria [54]. As a key contribution of this study, Cipollino aims to improve pitfalls of the previous AS-aware Tor client (Astoria) to achieve better security against network-level adversaries. Cipollino primarily depends on PathCache [59] to compute AS paths for its AS-aware path selection algorithm.

However, at the time of conducting this study in July 2016, PathCache [59] has not yet been made available for testing, and the source code of Cipollino has not been published.

## 2.9 Research motivations

Considering all of the aforementioned works, the threat of AS-level adversaries has been intensively studied for more than a decade. Unfortunately, none of the proposals has been successfully deployed and widely adopted by end users, particularly non-technical users, while the threat of AS-level adversaries still remains a long-standing problem. Moreover, most prior AS-aware Tor clients were published in the form of offline packets, which often consisted of a modified Tor client, and a snapshot of the Internet topology for users to run AS-path inferences themselves. There are two obvious drawbacks in this approach. First, it might be too complicated for non-technical Tor users to properly compile the source code, and to get the client updated when new versions of Tor Browser Bundle are released. Second, due to the dynamic nature of the Internet, the snapshot of Internet topology has to be updated continually to assure a near real-time AS path inference. Therefore, in this study, we introduce ASmoniTor to improve upon those drawbacks and plan to provide it to end users.

# Chapter 3 Anti-Raptor: Anti routing attack on privacy for a securer and scalable Tor

The proposal of anti-Raptor in this chapter is an investigation [60] conducted during the master's study of the author. It was accepted by the 17th IEEE International Conference on Advanced Communications Technology.

## 3.1 AS-level adversaries: problem revisited with real-time data analysis

At the time of this study, we re-analyzed the AS-diversity in the Tor network. By examining the consensus network status document retrieved at 12:00 UCT on March 27th, 2015, using Stem Python library [61], we found that although there were 6697 relays in the network, they were distributed among just 1220 unique ASes [1]. Moreover, a large portion of the total number of relays are associated with a small group of ASes, as shown in Table 3. In fact, more than 30% of bandwidth (KB/s) and nearly 25% of relays are allocated within the top 10 ASes, which may attract adversaries to exploit these ASes and try to sniff traffic traversing to/from them in order to conduct correlation attacks.

Table 3: The 10 ASes containing the greatest number of relays

| ASN | Cumulative BW | % BW | Total Unique Relays | % Relays |
|------|------|------|------|------|
| 16276 | 3592823 | 11.65 | 401 | 5.99 |
| 24940 | 2507641 | 8.13 | 302 | 4.51 |
| 3320 | 74010 | 0.24 | 194 | 2.90 |
| 7922 | 44446 | 0.14 | 178 | 2.66 |
| 12876 | 3125003 | 10.13 | 146 | 2.18 |
| 6830 | 59479 | 0.19 | 124 | 1.85 |
| 20473 | 121342 | 0.39 | 80 | 1.19 |
| 701 | 105370 | 0.34 | 78 | 1.16 |
| 36351 | 61310 | 0.20 | 77 | 1.15 |
| 12322 | 66531 | 0.22 | 68 | 1.02 |
| **Total** | 9757955 | 31.64 | 1648 | 24.61 |

---

[1] We map IP to AS using http://ipinfo.io/. The API uses MaxMind's datasets [76] to provide IP lookup service.

Recently, Sun *et al.* introduced RAPTOR [52], which is a suite of attack techniques that can be used by AS-level adversaries to compromise the anonymity of Tor users. The authors in [52] tested RAPTOR in the live Tor network and succeeded in compromising user anonymity (for ethical reasons, their experiments were tested within self-created Tor relays so that no actual user's privacy was compromised). In spite of successfully conducting real attacks on self-created relays, Sun *et al.* provided limited samples and self-created AS topology graphs to demonstrate their attacks, which does not reflect a more complex reality of the Tor network. Therefore, in order to gain a more comprehensive insight into the current threat posed by AS-level adversaries, we use real-time data collected from the Tor network to identify ASes at which a threat like RAPTOR could occur.

Table 4: Top 10 ASes that cover a large proportion of bandwidth

| ASN | Cumulative BW | % BW | Total Unique Relays | % Relays |
|---|---|---|---|---|
| 16276 | 3592823 | 11.65 | 401 | 5.99 |
| 12876 | 3125003 | 10.13 | 146 | 2.18 |
| 24940 | 2507641 | 8.13 | 302 | 4.51 |
| 8972 | 1568756 | 5.09 | 60 | 0.90 |
| 24961 | 1364022 | 4.42 | 54 | 0.81 |
| 60781 | 805707 | 2.61 | 57 | 0.85 |
| 37560 | 599144 | 1.94 | 3 | 0.04 |
| 43350 | 527741 | 1.71 | 10 | 0.15 |
| 51395 | 520938 | 1.69 | 22 | 0.33 |
| 13030 | 506128 | 1.64 | 19 | 0.28 |
| **Total** | **15117903** | **49.02** | **1074** | **16.04** |

Because previous works often focused on ASes containing the majority of relays, an analysis similar to the one presented in Table 3 (in which ASes are sorted by the number of relays they contain), might have been seen before. However, it is expected that a select group of ASes would have a relatively large view of the overall Tor network since the original Internet availability is unevenly distributed. Unlike other works, we also analyze the real-time data from Tor's consensus file and sort the ASes in Table 4 by cumulative relay bandwidth. Interestingly, there are 7 ASes that do not appear in Table 3, but that appear in Table 4 with high rankings. In other words, there is a paradox in which some ASes contain very few relays, but account for a striking amount of bandwidth of the overall network, and vice versa. With the top 3 ASes remaining from Table 3, the 10 ASes in Table 4 cover roughly 50% of the total bandwidth of the Tor network.

Owing to the recent bandwidth-based path selection algorithm, this group of ASes is an ideal target of AS-level adversaries, especially those ASes that do not contain a large number of relays but account for a significant amount of bandwidth. For example, AS37560 contains only 3 relays (46.246.46.27, 46.246.32.223, and 197.231.221.211), but occupies nearly 2% of the whole network's bandwidth, as highlighted in Table 4. A summary of these 3 relays is provided in Table 5.

Table 5:    Relays in AS37560

| IP | BW | Fast | Guard | Exit | Country | Multi-origin |
|---|---|---|---|---|---|---|
| 46.246.46.27 | 26000 | ● | ● | | SE | AS42708 |
| 46.246.32.223 | 144 | ● | | ● | SE | AS42708 |
| 197.231.221.211 | 573000 | ● | ● | ● | LR | None |

Although AS37560 was not listed in Table 3 and is only at rank #7 in Table 4, by hosting only 3 fast guard/exit relays, its probability of being selected as a guard or an exit is even higher than AS3320 (rank #3) and AS7922 (rank #4) in Table 3. We use Compass [62] to validate this conclusion. At the time of writing this paper, the overall probabilities of AS37560, AS3320 or AS7922 being chosen as a guard AS or an exit AS are 0.14%|2.63%, 0.02%|0.02% and 0.00%|0.06% respectively.

In this case, without any apparent evidence, we cannot intuitively conclude that AS37560, or any of its relays, is an adversary. However, due to Tor's current design (i.e., relays are selected by a bandwidth-based algorithm, and Tor anonymous circuits are created in a manner that relays in a circuit do not belong to the same /16 subnet), the ones who have control over particular ASes, such as ISPs or government agencies, would have no difficulty carrying out a correlation attack by following these steps:

1. Locating ASes which have distinctly fast guards/exits but do not host many relays (e.g., AS37560, AS43350)

2. Injecting another fast exit/guard, which does not belong to the same /16 subnet as the guard/exit detected in step 1, into the same AS

3. Snooping at AS-level to conduct the correlation attack with a lower cost since there are only few relays in the anonymity set hosted in that AS

Using these steps, the adversary does not even need to conduct BGP hijacking or intercepting attacks to monitor Tor traffic, as mentioned in RAPTOR [52].

Despite numerous proposals to configure the Tor client to pick up relays located in different countries [50] in order to avoid picking up IP addresses hosted under the same AS, our real-time data analysis shows that relays 46.246.46.27 and 46.246.32.223 are hosted in Sweden (SE) and relay 197.231.221.211 is hosted in Liberia (LR), regardless of operating under the same AS (AS37560). Therefore, the approach proposed in [50] is not effective enough to mitigate AS-level adversaries if a client accidentally uses one of the followings Tor paths:

- (46.246.46.27, any middle relay, 197.231.221.211)
- (197.231.221.211, any middle relay, 46.246.32.223)

Next, for the purpose of completeness, we keep using AS37560 as a real-life case to demonstrate how the interdependent and dynamic Internet topology (#2 RAPTOR attack fashion) and BGP intercepting (#3 RAPTOR attack fashion) could increase the ability of an AS to monitor more Tor traffic.

In order to visualize AS37560 using Internet topology, we investigated the latest dataset of AS relationships made available by CAIDA [63], which was retrieved on March 27th, 2015. We then obtained the graph in Figure 7.

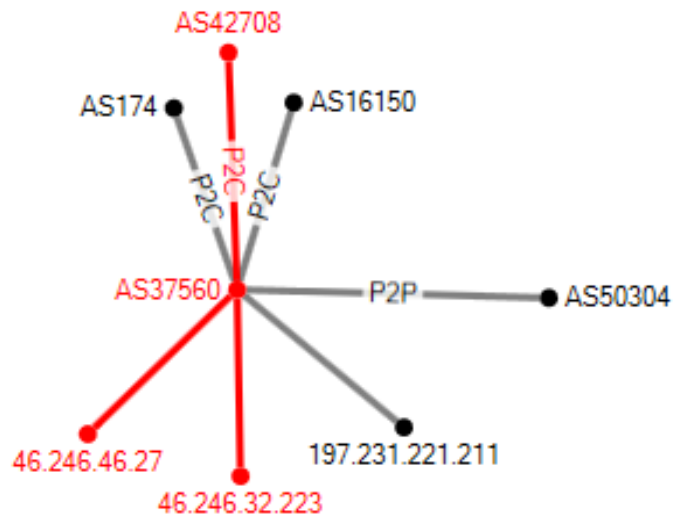

Figure 7: AS37560's Internet topology visualization

We found that AS37560 is a customer AS of 3 other provider ASes (AS174, AS42708 and AS16150) and has AS50304 as a peer AS. These connections indicate that the three provider ASes are the main gates for traffic traversing from/to AS37560 to/from the entire Internet. Consequently, the threat (if any) of a correlation attack posed by AS37560 is propagated to the 3 provider ASes, and the total probabilities of being chosen as guard/exit ASes should also be propagated because they also have the ability to monitor the traffic of AS37560. The recent design of Compass [62] does not take this circumstance into account. Therefore, we re-calculate the approximate probability of being chosen as a guard AS or an exit AS for these 3 ASes, as shown in Table 6. In terms of threat-propagated awareness, the probability of being chosen as a guard AS or an exit AS for each provider AS should be increased by $x$, with $x$ smaller than the guard probability and the exit probability of AS37560.

Table 6: Re-calculated probabilities of being chosen as a guard AS or an exit AS due to the interdependent nature of Internet topology

| AS | Number of Relays | Guard Probability | Exit Probability |
|---|---|---|---|
| 37560 | 3 | 0.14 | 2.63 |
| 174 | 9 | 0.73 + (x < 0.14) | 0 + (x < 2.63) |
| 42708 | 15 | 0 + (x < 0.14) | 0.02 + (x < 2.63) |
| 16150 | 0 | 0 + (x < 0.14) | 0 + (x < 2.63) |

Due to the complexity of Internet routing policies, and since traffic allocation between ASes is confidential information in contracts between ISPs, we cannot truly know how much traffic is allocated among these 3 provider ASes. In contrast, we can be sure that if one of these 3 ASes goes down due to infrastructure maintenance, corruption or a DDOS attack, the remaining two ASes will transmit more traffic of AS37560. If two of them are down, the remaining one will be able to monitor the entire traffic of AS37560. The interdependent and dynamic Internet topology could increase the ability of an AS to monitor more Tor traffic by this process.

Next, by mapping IPs to ASes in Table 5, using Hurricane Electric's BGP toolkit [64], we found that the first two IPs are announced by two ASes, AS37560 and AS42708, as indicated in the last column of Table 6. However, only AS37560, with a more specific advertised prefix (i.e., 46.246.32.0/**19**) is retrieved by https://ipinfo.io/ in Tables 3 and Table 4. AS42708, with a less specific advertised prefix (i.e., 46.246.0.0/**17**), is not retrieved. We also found that AS42708 is a provider AS of AS37560, as illustrated in Figure 7.

At the time of retrieving data from Compass on March 27th, 2015, there was no Tor relay operating under AS16150. However, as discussed in RAPTOR attack fashion #3, AS16150 can still monitor Tor traffic by maliciously advertising a more specific prefix than AS42708, such as 46.246.0.0/**18**. In this case, the traffic is still redirected to its correct destination, so AS16150 would successfully snoop on Tor's traffic traversing from/to the entire Internet to/from AS37560. BGP hijacking can also be conducted in a similar manner with BGP intercepting, but it would not last for a long time since hijacking BGP is just a black hole that does not send response packets to the original sender, thus communication would eventually drop. That is a scenario in which an AS-level adversary can abuse a BGP intercepting attack to increase its ability to sniff Tor traffic.

## 3.2 Design goals of anti-Raptor

We propose anti-Raptor as a countermeasure to the AS-level attack. Since we showed in section 1.2.3 that an increase in number of clients results in a higher bandwidth cost of serving the consensus document, and because the recent scalability problem has not been efficiently handled, our proposal is designed in a manner to avoid introducing unnecessary load to the network. In particular, load that is directly related to the number of clients, which will likely result in additional scalability problems.

Although RAPTOR is referred to in this study as the newest attack technique on privacy in Tor and it was first published on March 13th, 2015, our proposal is not merely a countermeasure to this type of attack. Instead, it has been studied and designed as a long term solution after Edward Snowden confirmed the NSA has an autonomous system called MonsterMind [65], which is said to be able to intercept network traffic flows. Therefore, our proposal is not just a reaction to RAPTOR, but rather a long-term carefully planned countermeasure to mitigate the threat of AS-level adversaries in general.

### 3.3 System architecture

First of all, it is important to note that we cannot dictate where volunteers should run their relays for two primary reasons. First, Tor is operated by a network of volunteer-based relays that are not evenly distributed on the Internet, since the Internet availability is not evenly distributed around the globe. Second, different ISPs and governments have different Internet policies. As a result, mitigating real-world threats of AS-level adversaries, like the case of AS37560 discussed in section 3.1, is not a trivial task.

To negate the threat of AS-level adversaries, we suggest that Tor clients initiate Tor paths in an AS-aware manner in which AS paths between the client and the guard, as well as between the exit and the destination, are measured.

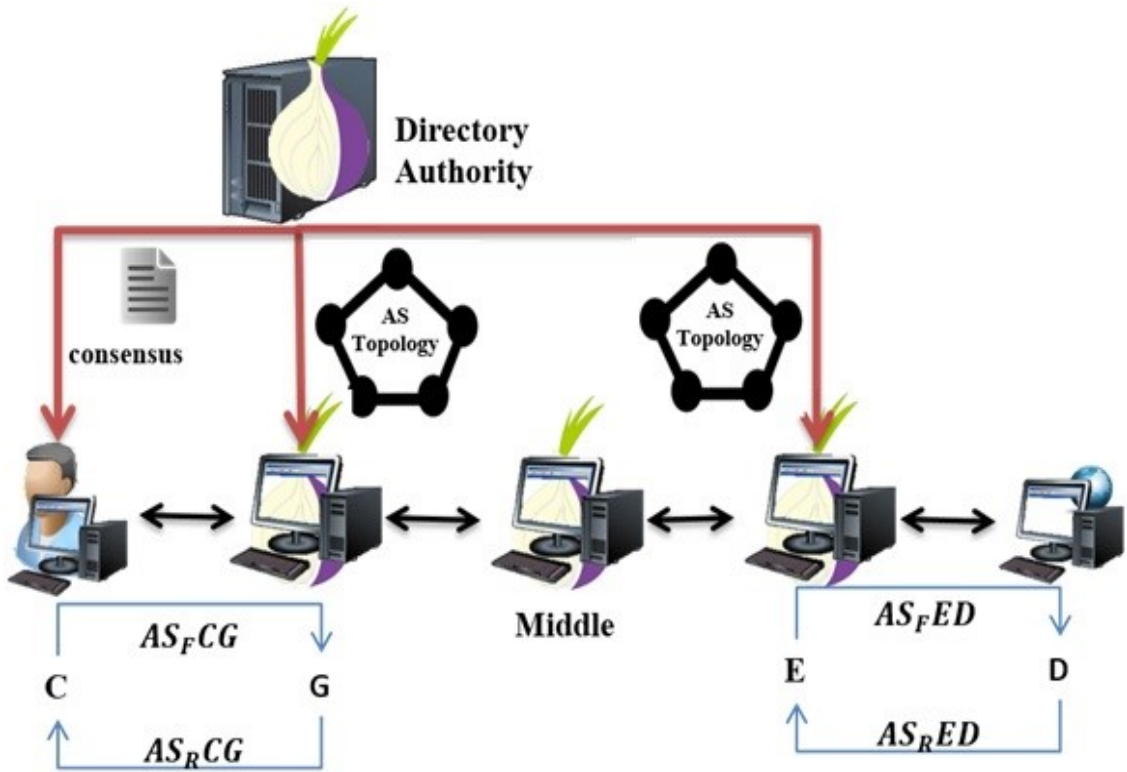

Figure 8: Design of anti-Raptor

- $AS_F CG$: ASes set in forward link between client and guard
- $AS_R CG$: ASes set in reverse link between client and guard
- $AS_F ED$: ASes set in forward link between exit and destination
- $AS_R ED$: ASes set in reverse link between exit and destination

### 3.3.1 AS-awareness

In anti-Raptor, we suggest that an AS topology is maintained and distributed to guard and exit relays by trusted central directory authority servers. It is a directed graph of the Internet AS-level topology that will be used by guard and exit relays to look up four AS paths: $AS_FCG$, $AS_RCG$, $AS_FED$, and $AS_RED$. In the current Tor design, anonymous circuits are built based on the bandwidth-preferred algorithm, and relays are selected so that they do not belong to the same /16 subnet. However, as shown in section 3.1, these constraints do not effectively and efficiently mitigate AS-level adversaries. Therefore, clients should pick up circuit in an AS-aware manner to avoid using circuits in which the same AS appears at both ends of the communication. Hence, before initiating a particular circuit, the client sends two requests to check AS paths between herself and the guard, and between the exit and the destination at the same time she sends requests to open the virtual circuit. After receiving requests, the guard and the exit look up the AS topology file to measure $AS_FCG$, $AS_RCG$, $AS_FED$, and $AS_RED$, sending the results to the client. Based on the measured ASes sets received from the guard and the exit, the client implements this algorithm:

| AS-awareness algorithm |
|---|
| **if**: |
|    $(AS_FCG \cup AS_RCG) \cap (AS_FED \cup AS_RED) = \emptyset$ |
| **then** initiate circuit |
| **else** drop |
| go to next circuit in the list of pre-emptively built circuits |

### 3.3.2 Scalability

Our approach is similar to LASTor [50] in term of the AS-path measuring procedure. However, LASTor requires Tor users to initially download 13 MB of AS-related data, including inter-AS links, AS three-tuples, and AS path lengths. We already knew that AS topology is very dynamic, thus maintaining such files from the client side in order to have near real-time AS-path inferences is not easy, and it may cause more loading overhead in the Tor network. On the contrary, our design only requires guard and exit relays to download the AS topology, instead of all clients. The AS topology only needs to be updated when a new one is published by directory authority servers. One may question whether the size of the response packets, containing AS-path measurements, from guard and exit relays to the client, would add more load to the Tor network. However, we already analyzed the AS topology from CAIDA [63] and found that average paths have only 4 ASes. Based on that information, response packets containing AS numbers will not cause serious loading overhead to the Tor network. Additionally, to preserve the recent Tor's approach of improving performance by using a bandwidth-based algorithm, we apply anti-Raptor directly on the top of the list of pre-emptively built circuits. From top to bottom (i.e., from fast circuits to slow circuits), all circuits are checked before being initiated.

## 3.4 Discussions

In a compatible manner with the proposal of Dingledine *et al.* [66], if the guard is stable and used for a long period of time in future Tor design, then cost for computing ($AS_F CG \cup AS_R CG$)can be reduced, since the client only needs to request the guard to report ($AS_F CG \cup AS_R CG$) for the first time and reuses the result until a reasonable time has passed (up to 9 months in future design [66]).

In [52], the authors suggest that the user should choose a closer guard relay to lower the possibility that an adversarial AS appears in the channel between herself and the guard. However, such an approach may make the user's traffic distinguishable or cause traffic bottlenecking issues. Thus, the approach of shorter-Tor-path AS still needs to be carefully researched and will be discussed briefly in Chapter 4. As we found that the average distance between two arbitrary ASes is 4, future work may use this number to estimate how many ASes a path should have to negate the threat of AS-level adversaries. In addition, depending on the length of the AS path, one can decide after how long the ($AS_F CG \cup AS_R CG$) should be re-measured. The shorted AS paths are, the less dynamic they become. The authors in [52] also recommended that Tor advocate its voluntary relay operators to run Tor relays with a prefix longer than /24 to alleviate BGP hijacking and intercepting attacks. However, as mentioned above, it is not easy to dictate where volunteers should run their relays. Furthermore, BGP hijacking and BGP interception occur when an AS tries to advertise a more specific prefix for various ill intensions. In other words, when this kind of attack occurs, the entire Internet would likely see two different ASes announce a common range of prefixes. However, this assumption is not always true because while multi-origin routes can be an indication of BGP hijacking, they can also occur for other reasons. For example, they could occur in case of an organization not having their own AS, thus requesting many ISPs to announce their IP space for them, or in the case of anycast usage from various ASes. By observing a BGP toolkit provided by Hurricane Electric Internet service [64], we found that daily updated multi-origin routes include 6600 unique prefixes, half of which are /24 subnets, advertised by 14000 different ASes. Therefore, if we drop any circuit based solely on the AS testing result, without carefully taking this circumstance into account, we would waste Tor's limited resources, since the phenomenon of multi-origin routes occurs frequently. To that end, we suggest that trusted central directory authority servers also observe the multi-origin routes and blacklist those ASes that advertise the same prefix for the first time, and remove them from this blacklist after a "trustable" period of time has passed. The list of blacklisted ASes should be distributed together with consensus (it should not be distributed by default, but only to those clients who opt for an AS-aware path selection algorithm). Additionally, we can take complete advantage of bandwidth scanning authority servers to detect abnormal changes in the Internet AS-level topology, while sending packets to guard and exit relays to test their bandwidth.

Finally, it is worth noting that a threat to our design only occurs when both guard and exit relays try to strategically, systematically, and simultaneously lie about the AS paths of $AS_F CG$, $AS_R CG$, $AS_F ED$, and $AS_R ED$, preventing the client from recognizing that she is going to use a circuit in which the same adversarial AS will appear at both ends. Nevertheless, thanks to the conventional wisdom of the three-relay structure of a typical Tor circuit, the guard cannot know which exit is at the opposite side of the circuit, and vice versa. A situation in which both guard and exit relays know that they are handling the same circuit and provide false responses about AS paths is unlikely to occur. Integrating anti-Raptor proposal into the current Tor network would require major restructuring; therefore, we will introduce ASmoniTor in Chapter 5, which is a stand-alone platform that operates based on the anti-Raptor algorithm.

# Chapter 4 Shorter-Tor-Path as a Solution

As briefly mentioned in previous studies [49], [50], [52], a shorter AS path can effectively help to reduce the risk of an adversarial AS being able to compromise the anonymity of Tor users through correlation or RAPTOR attacks. Although such an approach is promising, no previous works have examined its feasibility. In this chapter, we examine the possibility of implementing this approach by studying the potential distributions of censored Internet users, guard nodes, exit nodes, and most likely visited websites via Tor. By visualizing these distributions on a map, we preliminarily examine the possibility of implementing this approach.

## 4.1 Premise

First of all, it is necessary to define what a shorter Tor path is. Since the purpose of this study is to find a better method to help Tor users negate the risk of being deanonymized by AS-level adversaries, we are interested in shortening the routes between the two ends of anonymous Tor paths. In other words, we aim to shorten routes between (1) the client and the guard, (2) the exit and the intended destination host, and (3) both (1) and (2) at the same time, which is the most desired target. In this context, a shorter Tor path should not be confused with a shorter Tor path from the guard node, via the middle node, to the exit node, because the attack targets of AS-level adversaries are the two ends of a Tor path, but not the middle node. Therefore, by shortening the routes at both ends of a Tor path, we can reduce the probability that an adversarial AS appears at the two ends of that Tor anonymous circuit.

## 4.2 Modeling the involved parties

In order to determine whether or not the aforementioned approach is possible, we have to know where Tor clients, guard nodes, exit nodes, and the most likely visited websites are located. The shorter-Tor-path approach can only be realized if clients are located close to guards, and exits are close to destination hosts. In order to be considered "close", the route between two entities must be short in the concept of network distance, but not necessarily physical distance.

Although an IP address is not always tied to a physical location and its GeoIP

information may be incorrect at times. The distribution of IP addresses and Internet users provides a good approximation of the distribution of the human population. Similarly, Internet backbones and data centers are designed and located in accordance with the distribution of the human population to facilitate economical routing.

To that end, we employ the GeoIP data to map IP addresses according to their physical locations (i.e., longitude and latitude), with the hope of gaining some insight into the distribution of Tor users, guard nodes, exit nodes and most likely visited websites. Finally, we present these distributions on a map in the following subsections.

### 4.2.1 Where are the clients?

As described in Figure 1, there are only two entities in the Tor network that can obtain IP addresses of Tor users: directory servers and guard nodes. Tor users need to download the consensus document from the directory authority server to maintain an up-to-date view of the Tor network, while guard nodes are the entrances for directly connecting Tor users to enter the Tor network. Therefore, one of the most effective ways to obtain IP addresses of Tor users is to create many fast and stable guard nodes, and to enable them to become mirror servers of the directory servers. Using this method, we can obtain and log a sample of Tor user IP addresses, the locations of which likely reflect the real distribution of Tor users.

However, with the issue of ethics regarding the collection and publication of Tor users' real IP addresses [58], [67], we opt to reuse the analyzed results made available by Tor Metrics project [24]. The following is a table of the top 10 countries by number of directly connecting users, which make up nearly 65% of the total number of Tor direct daily users.

Table 7: Top 10 countries by directly connecting users

| | Country | Mean daily users | | Country | Mean daily users |
|---|---|---|---|---|---|
| 1 | United States | 359432 (19.88 %) | 6 | Italy | 49869 (2.76 %) |
| 2 | Russia | 206737 (11.43 %) | 7 | Spain | 49053 (2.71 %) |
| 3 | Germany | 177034 (9.79 %) | 8 | Brazil | 48945 (2.71 %) |
| 4 | France | 108474 (6.00 %) | 9 | Japan | 47375 (2.62 %) |
| 5 | United Kingdom | 82582 (4.57 %) | 10 | Canada | 39653 (2.19 %) |

### 4.2.2 Where are the guards and exits?

The Tor consensus file was fetched at 16:30 UCT on July 9th, 2016, and analyzed to determine the location of most guard and exit nodes. In total, we were able to retrieve 2277 guard nodes and 850 exit nodes, which were then mapped using GeoIP information. The distribution of these nodes is presented in Figure 9 and Figure 10.

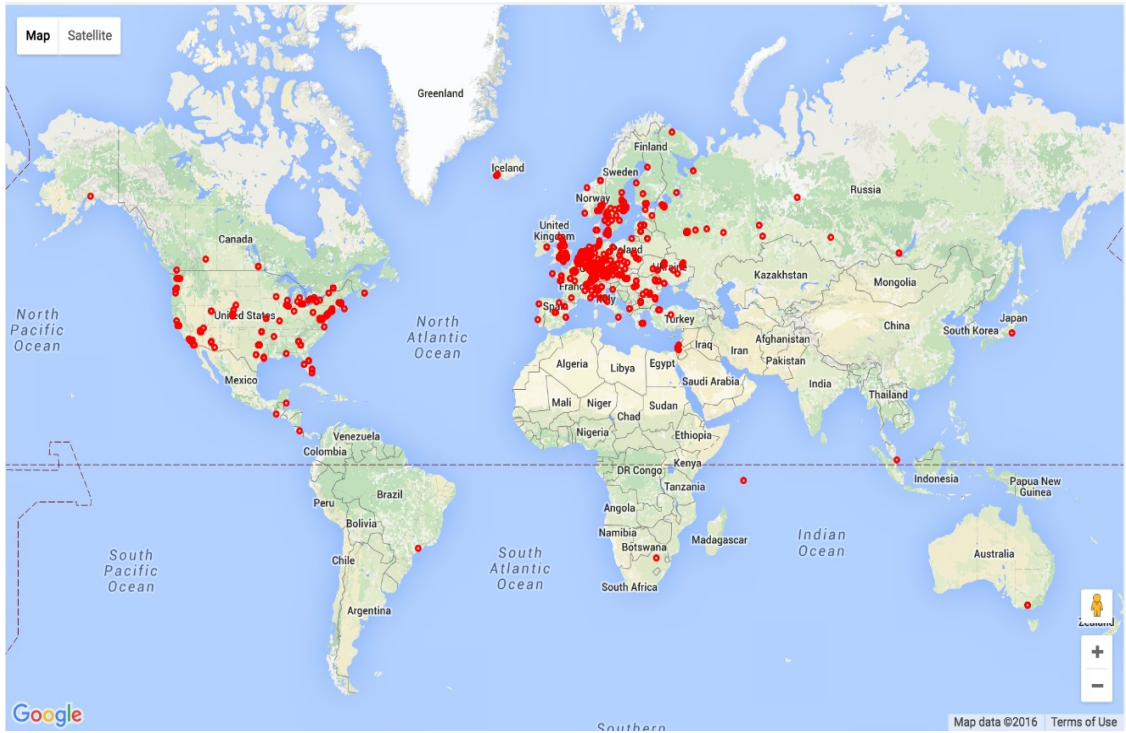

Figure 9: Distribution of guard nodes

As visualized on the map, most guard nodes are hosted in Europe and the United States. Table 8 shows the analyzed results of the top 10 countries hosting the largest number of guard nodes.

Table 8: Top 10 countries hosting the largest number of guard nodes

|   | Country | Guard | Percentage |    | Country | Guard | Percentage |
|---|---------|-------|------------|----|---------|-------|------------|
| 1 | FR      | 471   | 20.69%     | 6  | SE      | 100   | 4.39%      |
| 2 | DE      | 422   | 18.53%     | 7  | CA      | 97    | 4.26%      |
| 3 | US      | 298   | 13.09%     | 8  | RU      | 75    | 3.29%      |
| 4 | NL      | 204   | 8.96%      | 9  | CH      | 67    | 2.94%      |
| 5 | GB      | 112   | 4.92%      | 10 | RO      | 50    | 2.20%      |

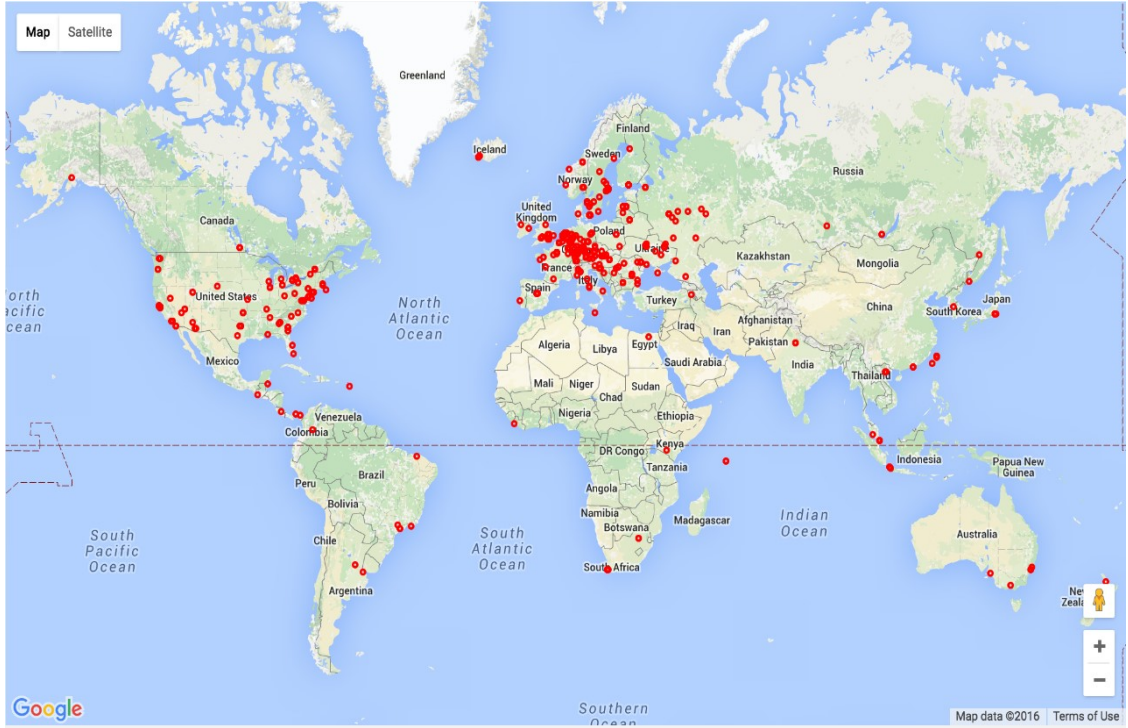

Figure 10: Distribution of exit nodes

Similar to the distribution of guard nodes, exit nodes are also mostly located in Europe and the United States. However, the number of exit nodes is only about 1/3 the number of guard nodes. Table 9 shows the summarized result of the top 10 countries where exit nodes are hosted.

Table 9: Top 10 countries hosting the largest number of exit nodes[1]

|   | Country | Exit | Percentage |    | Country | Exit | Percentage |
|---|---------|------|------------|----|---------|------|------------|
| 1 | US      | 140  | 15.42%     | 6  | CA      | 45   | 4.96%      |
| 2 | NL      | 110  | 12.11%     | 7  | RO      | 43   | 4.74%      |
| 3 | FR      | 83   | 9.14%      | 8  | SE      | 39   | 4.30%      |
| 4 | DE      | 79   | 8.70%      | 9  | RU      | 37   | 4.07%      |
| 5 | GB      | 51   | 5.62%      | 10 | CH      | 31   | 3.41%      |

As a result, we can see that although rankings differ slightly between Table 8 and Table 9, the 10 countries hosting the largest number of both guard nodes and exit nodes are the same.

---

[1] Countries in Table 8 and 9 are noted by ISO 3166-1 alpha-2 codes (two-letter country codes).

### 4.2.3 Where are potential intended destination hosts?

Perhaps the easiest way to acquire IP addresses of websites, which are often visited by Tor users, is to create a large number of exit nodes, distributed in different locations, and monitor traffic from those exit nodes to the visited destinations. Nevertheless, due to the anonymous nature of the Tor network, and because of ethical concerns with logging and publishing websites visited by Tor users, it is not a trivial task to obtain a complete list of the most common intended destination hosts of Tor users. This type of method should not be employed, as was discussed in section 2.2.

Since doing experiments on the live the Tor network directly is not recommended, some previous works determined the most visited websites using author opinion[48], or third party statistics of the most visited websites on the Internet [50], [54]. The authors in [50] modeled the destination hosts using the top 200 most visited websites in the United States, as determined by Quantcast [68], while the authors in [54] used the top 100 most visited websites, as analyzed by Alexa, along with 100 country-specific sensitive webpages collected by the Citizen Lab [69].

However, after careful examination of the websites used in the studies described above, we argue that the most visited websites in the normal Internet environment are not the same as the most visited websites via the Tor network. For instance, in the top site lists obtained from both Quantcast and Alexa[1], there are many shopping (e.g., Amazon, eBay, Rakuten, Alibaba, etc.) and electronic wallet (e.g., PayPal) websites. Theses websites know the identity of their users, even when they are accessed via Tor, thus it is not necessary to use Tor to access such webpages. The authors in [54] used a better method to model intended websites by integrating 100 country-specific topic-sensitive webpages collected by the Citizen Lab. However, this list of websites was assembled by GlobalVoices Blockpage Gallery [70], which states on their homepage that the list was generated using data from 2008 and does not reflect current online censorship.

Furthermore, the GlobalVoices Blockpage Gallery project collects blocked

---

[1] A list of the top 100 websites obtained from Alexa on July 13th, 2016, is provided in the Appendix.

websites reported by Internet users, while state-sponsored surveillance activities are not limited to blocking websites. Sometimes, repressive regimes cannot completely prevent their citizens from accessing censored websites due to the rapid development of pro-privacy and anti-censorship tools. Instead, they allow citizens to freely access censored websites, but secretly monitor their traffic to put them into the government's watch list, or even black list.

Taking into account of all the aforementioned issues, we can conclude that none of the prior studies' modeling methods are reliable enough to accurately reflect the most visited websites via Tor Browser, while knowing the correct destination hosts plays a very important role in measuring the risk posed by AS-level adversaries.

To improve upon the pitfalls of previous studies, we carefully review the annual survey of Freedom on the Net [71], conducted by the Freedom House, to examine which content is often monitored, censored, or blocked on the Internet. According to the report, the most censored topics include those that criticize or oppose the authorities, political opposition, satire, social commentary, adult-oriented material, and GLBT-related content. Fortunately, in addition to analyzing the most popular websites, Alexa also clusters them into categories similar to the above censored topics. Because of the similarity between the clustered websites and censored topics on the Internet, we fetched a list of these categories[1] and displayed them on map, as shown in Figures 12 through 18.

Comparing the maps displayed in Figures 12 through 18 with the map shown in Figure 11 reveals a visible difference between the distribution of topic-oriented destination hosts and the distribution of the top 100 websites. While non-categorized top 100 websites are evenly distributed throughout North America, European countries, and Northeast Asia, the categorized websites are mostly hosted in the United States and Europe. This comparison proves that using the non-categorized websites is inappropriate for modeling the websites likely visited by Tor users because it may yield unreliable results when measuring the threat of AS-level adversaries.

---

[1] We also append these lists of categorized websites obtained from Alexa on July 13th, 2016 at the end of the thesis in Appendix Section.

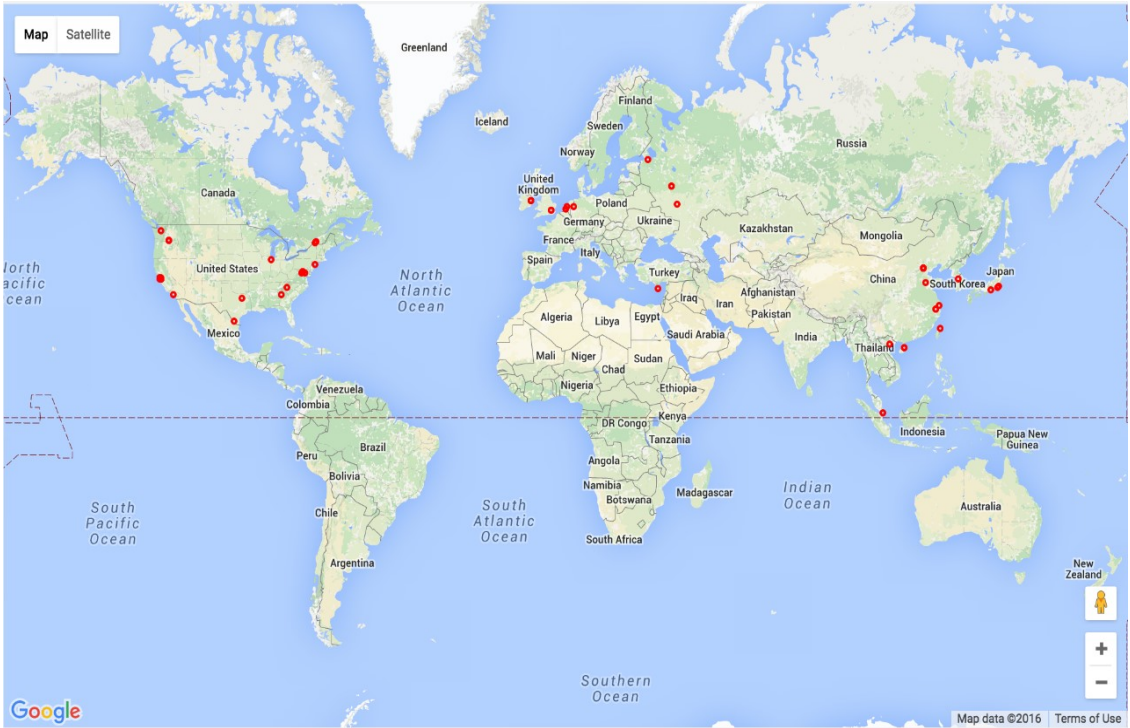

Figure 11: Top 100 websites analyzed by Alexa

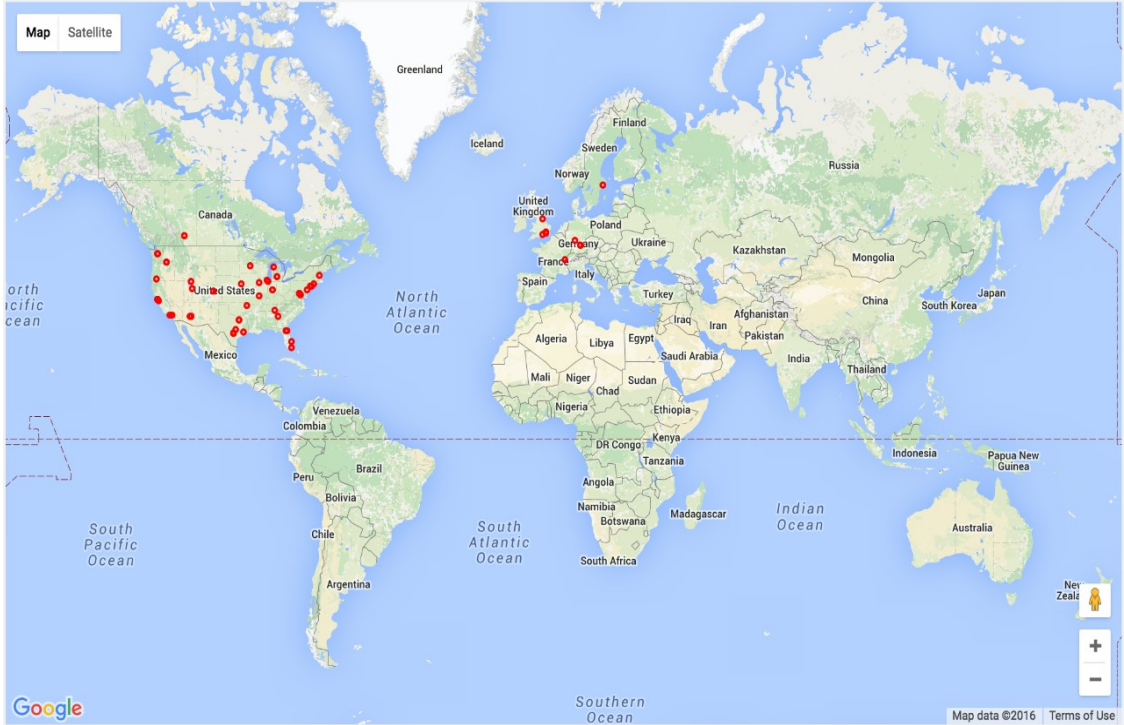

Figure 12: Top 100 websites in the politics category

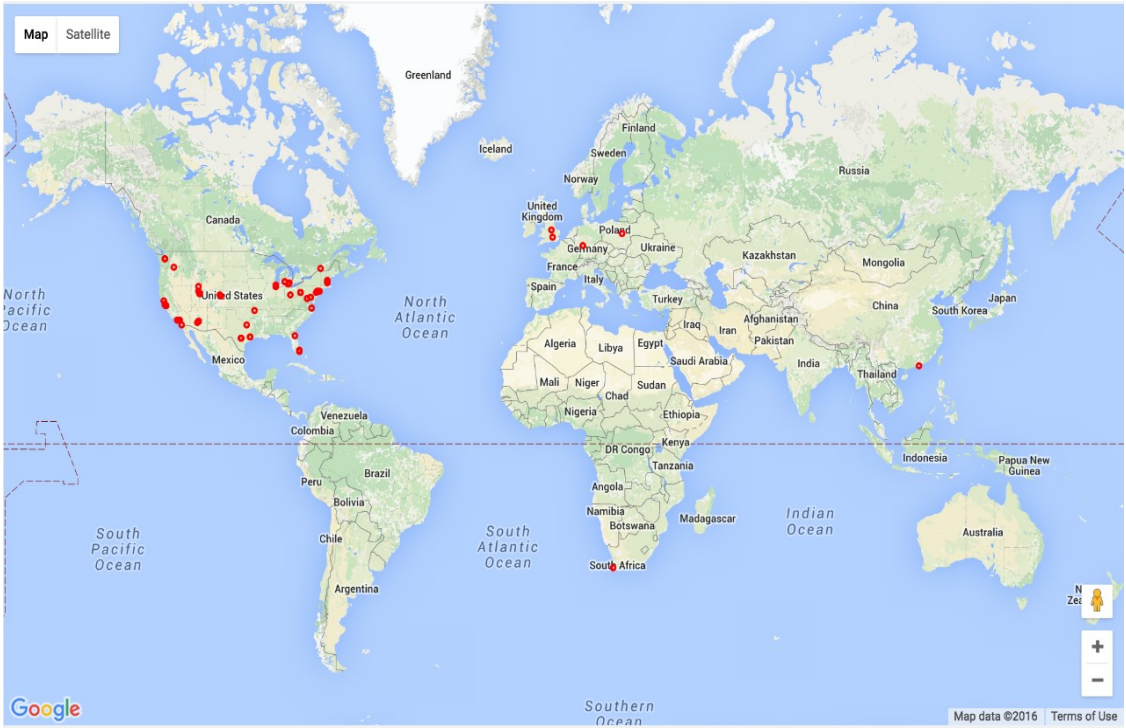

Figure 13: Top 100 in the "criticism and opposition of authorities" category

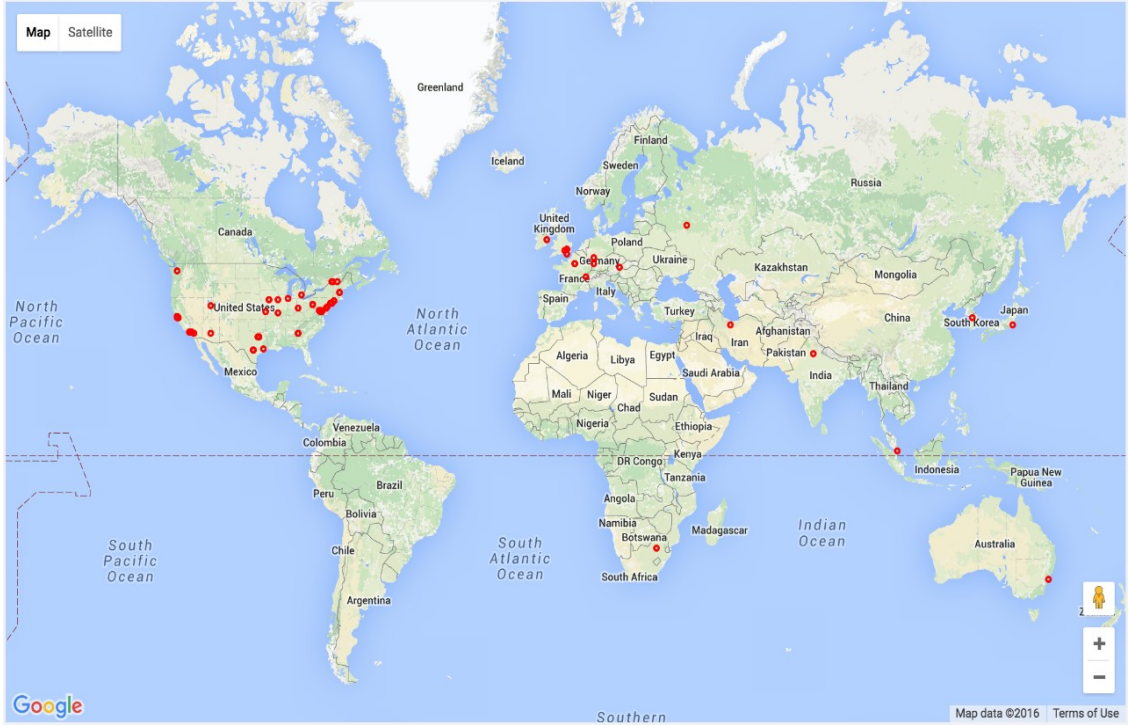

Figure 14: Top 100 websites in the media industry category

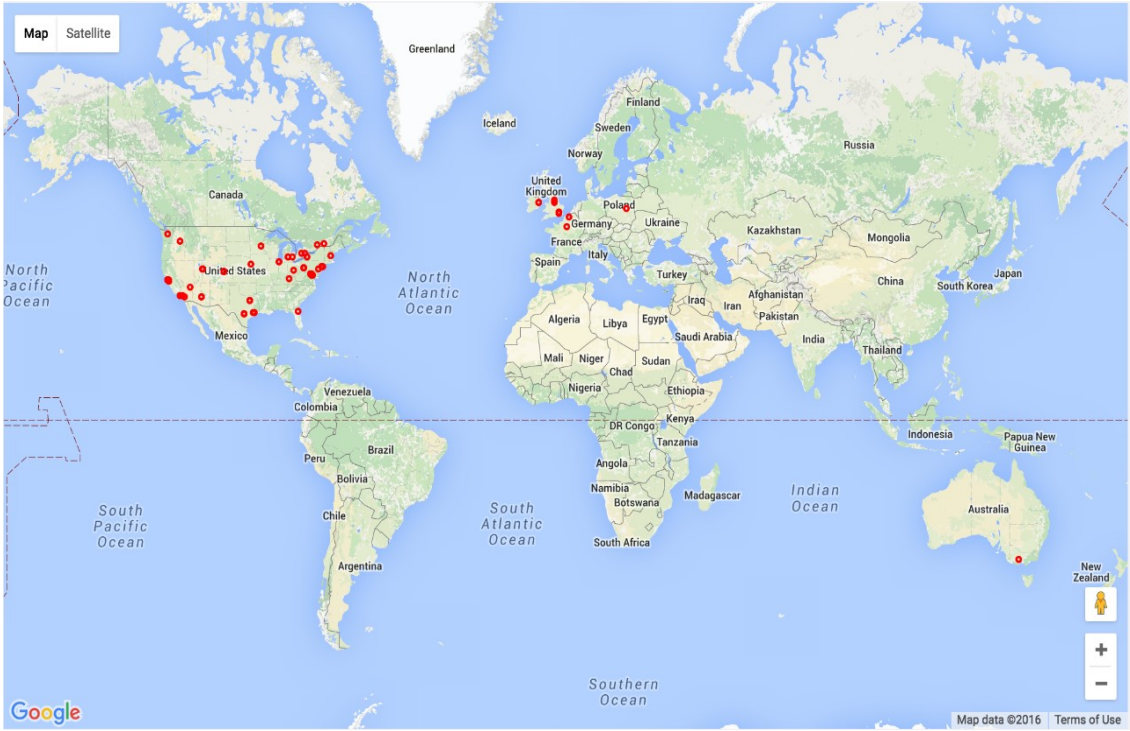

Figure 15: Top 100 websites in the activism category

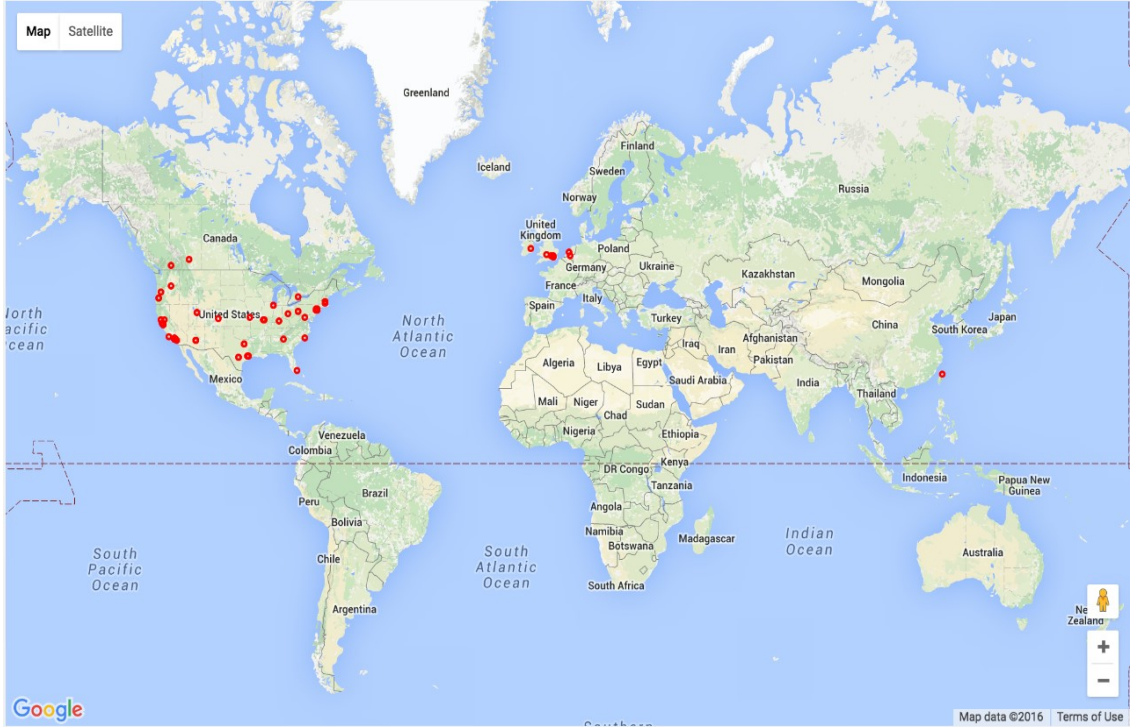

Figure 16: Top 100 websites in the support group category

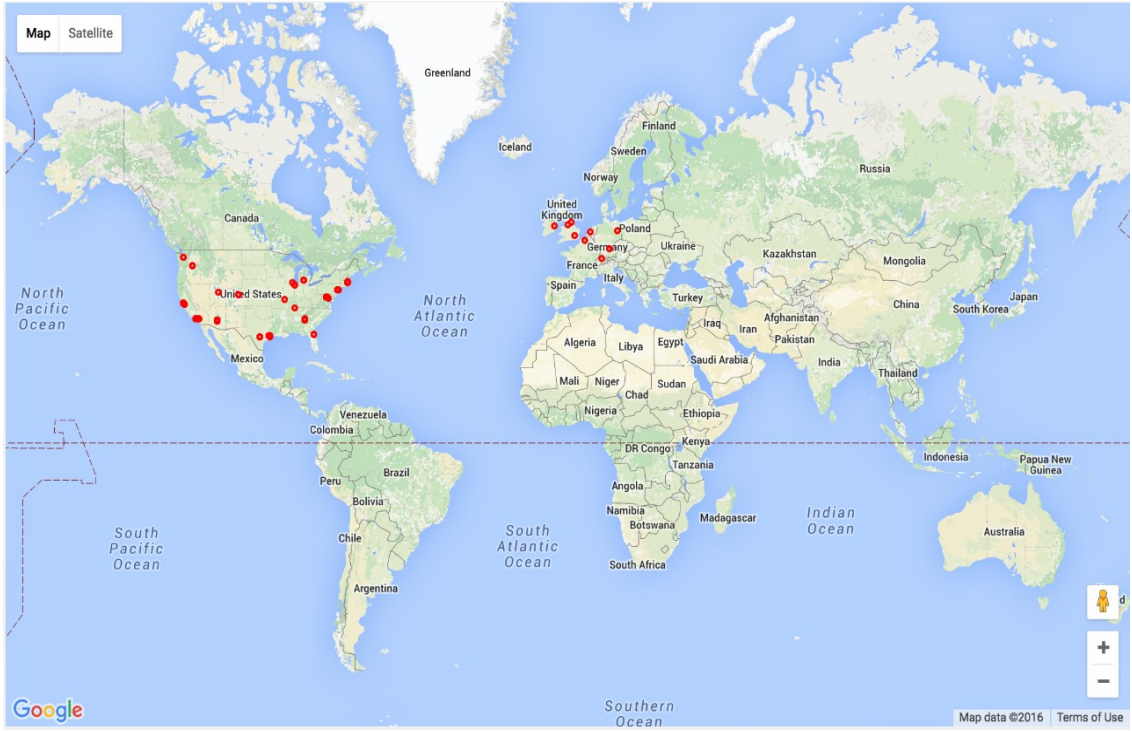

Figure 17: Top 100 websites in the GLBT category

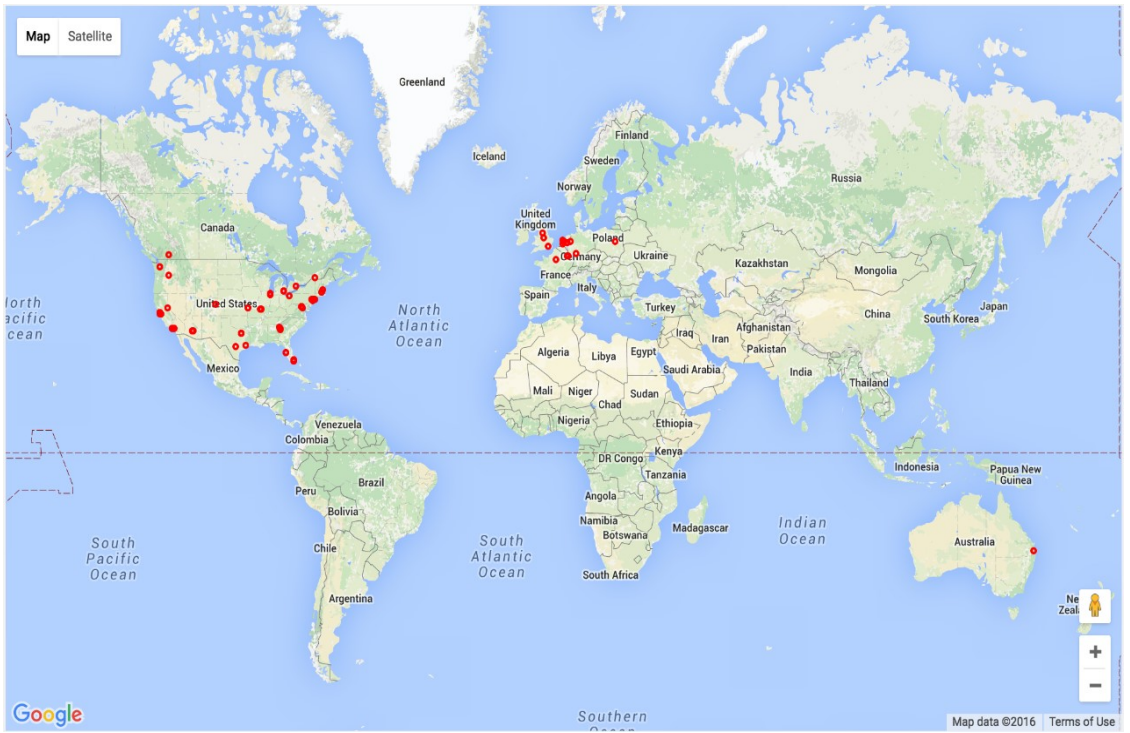

Figure 18: Top 100 websites in the adult category

### 4.2.4 Putting things together

**Can we shorten routes between users and guard nodes?**

By doing some calculations between Table 7 and Table 8, we can conclude that the approach of shortening routes between Tor users and guard nodes is feasible in some countries. In other countries, this approach needs further research to ensure that it will not cause a traffic bottleneck at guard nodes. As shown in Table 10, we calculated the ratio of direct users/one guard node to demonstrate this conclusion.

Table 10: Ratio of direct users/one guard node in the top 10 countries by number of directly connecting users

| Country | Mean daily users | Guard | Ratio | Feasibility |
|---------|------------------|-------|-------|-------------|
| France | 108474 | 471 | 230 | Possible |
| Canada | 39653 | 97 | 409 | |
| Germany | 177034 | 422 | 420 | |
| United Kingdom | 82582 | 112 | 737 | |
| United States | 359432 | 298 | 1206 | |
| Italy | 49869 | 32 | 1558 | |
| Russia | 206737 | 75 | 2756 | |
| Spain | 49053 | 8 | 6132 | |
| Japan | 47375 | 1 | 47375 | |
| Brazil | 48945 | 1 | 48945 | Impossible |

By shortening routes between users and guard nodes, we attempted to keep traffic entering the Tor network as local as possible to avoid having to traverse via many ASes, thus reducing the probability of an adversarial AS appearing. Currently, there are around 2 million daily users and around 2000 guard nodes in the Tor network, so we can approximate that the average number of users handled by each guard node is 1000. Basing on this result and the ratios in Table 7, it is possible to localize the entering Tor route in France, Canada, Germany and the UK, but it is difficult to implement this approach for Tor users in the United States, Italy, and Russia. It is nearly impossible to implement this approach for users in Spain, Japan and Brazil, since guard node(s) in these countries would be saturated by the large number of connections and circuits if many users tried to connect to Tor at once.

**Can we shorten routes between exits and intended destination hosts?**

As for shortening the outgoing Tor route from exit nodes to intended destination hosts, we overlap Figure 10 with Figures 13 through 18 to evaluate the approach's feasibility. Our findings indicate it is feasible to implement the approach of shortening the outgoing route, since most exit nodes are located in European and the United States, while most intended websites are also hosted in North America and Europe.

## 4.3 Even shorter paths cannot help: a remaining dilemma
### 4.3.1 When is Tor useless?

Although Tor is one of the most popular tools used to protect the anonymity of Internet users, there is a situation in which at least one AS-level adversary can deanonymize the Tor user, something we will call the "loop phenomenon". It occurs when a client uses Tor to visit a website that is also hosted in the same AS as her ISP. In this situation, incoming traffic from the client to the guard node has to pass through the AS of the user's ISP, regardless of the guard node's location. At the same time, outgoing traffic from the exit node to the intended website server also has to pass through the AS of the client's ISP. As a consequence, the ISP of the Tor user possesses an AS that is in an ideal position to carry out a correlation attack. This phenomenon is also likely to occur when the user visits a website hosted in her home country (i.e., the website is not necessarily hosted in the same ISP of the user). The following case illustrates the above scenario. For this example, a Tor user (in this case, my own PC, located in JP) accessed http://english.kyodonews.jp using Tor browser. Detail of the client, guard, exit nodes, and destination host is presented in Table 11.

Table 11: A case study of Tor's dilemma[1]

| Role | Global IP address | AS Number | Country |
|------|-------------------|-----------|---------|
| Tor client | 119.107.206.177 | AS2516 | Japan |
| Guard node | 23.239.10.144 | AS63949 | US |
| Exit node | 77.247.181.163 | AS43350 | Netherlands |
| Destination host | 202.248.158.165 | AS2510 | Japan |

[1]  We retrieve ASN and country of an IP using http://ipinfo.io/.

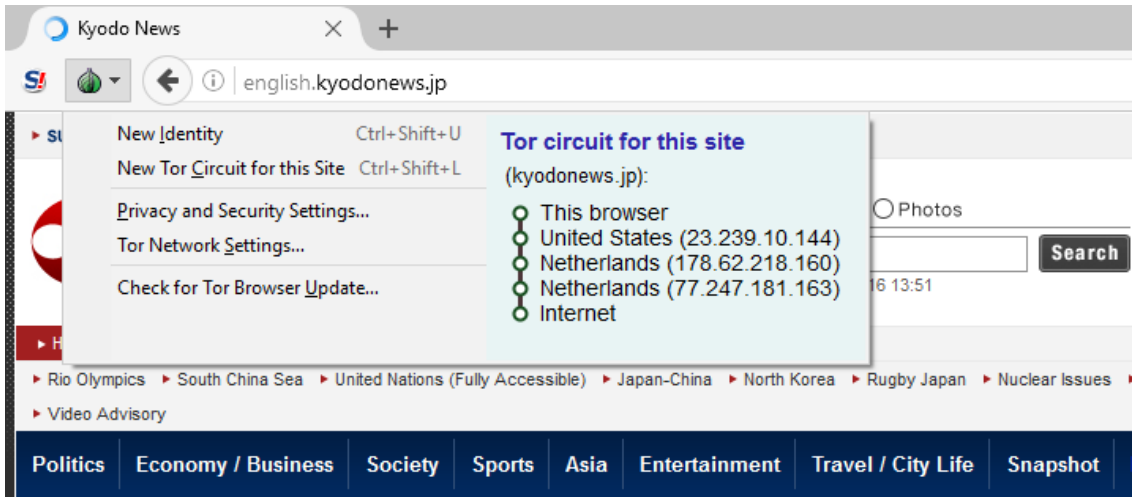

Figure 19: Accessing Kyodo News via Tor browser

To prove the statement above, we determined the set of ASes appearing between the two ends of the Tor circuit shown in Figure 19. They include the ASes that appear between the client PC (119.107.206.177) and the guard node (23.239.10.144), as well as the ASes that appear between the exit node (77.247.181.163) and the server of Kyodo News (202.248.158.165). When appearing at both ends of the Tor path, an AS is in an ideal position to carry out a correlation attack to deanonymize the client.

There are two components often used to measure the routing information of an AS path between two points on the Internet: control-plane and data-plane. Control-plane is used for measurements in which we have physical control over the computational resource. For instance, when applied to traceroute measurement, control-plane yields a reliable result since it is obtained in a real-time fashion. On the other hand, although less reliable, data-plane is utilized when we do not have physical access to the resource and must rely on known data to infer the routing information of the AS path.

We ran a traceroute measurement from the client PC to the guard node to get a real-time AS path that packets traverse from the client to the guard node. The AS path inference technique (data-plane) was used to find out the ASes that the route traverses from the exit node to the server of Kyodo News because we did not have physical access to the exit node.

```
PC-A385-Phong: ~ admin$ traceroute -a 23.239.10.144\
traceroute to 23.239.10.144, 64 hops max, 52 byte packets\
1 [AS56220] aterm.me (192.168.0.1) 3.423 ms 1.904 ms 1.747 ms\
4 [AS2516] 119.107.206.177 (119.107.206.177) 95.552 ms 112.055 ms\
5 [AS2516] 111.87.11.9 (111.87.11.9) 87.053 ms 92.328 ms 77.748 ms\
6 [AS2516] obpjbb205.int-gw.kddi.ne.jp (113.157.227.17) 102.017 ms\
7 [AS2516] pajbb002.int-gw.kddi.ne.jp (203.181.100.38) 206.227 ms\
8 [AS2516] ix-pa9.int-gw.kddi.ne.jp (111.87.3.58) 233.947 ms\
9 [AS3257] ae12.pao10.ip4.gtt.net (77.67.77.25) 205.316 ms 208.969 ms\
10 [AS3257] xe-2-1-7.nyc20.ip4.gtt.net (89.149.133.234) 1962.607 ms\
11 [AS3257] ip4.gtt.net (173.205.45.210) 1651.264 ms 1673.371 ms\
12 [AS8001] 209.123.10.77 (209.123.10.77) 1343.765 ms 276.309 ms\
13 [AS8001] 207.99.112.130 (207.99.112.130) 267.593 ms 249.936 ms\
14 [AS63949] switch-nacspare.linode.com (173.255.239.5) 248.076 ms\
15 [AS63949] tor.shamm.as (23.239.10.144) 273.352 ms 1468.360 ms\
16 [AS63949] tor.shamm.as (23.239.10.144) 1781.215 ms 274.452 ms\
```

Figure 20: Traceroute result from the client to the guard node

Figure 21: Result of AS path inference from exit node to Kyodo News[1]

---

[1] Note that the AS inference technique returned a list of AS paths in descending order of the probability that the route between two points would traverse on the real Internet. Packets sent from an IP address originating from AS43350 would mostly traverse via AS3257 and AS2516 before arriving at AS2510, while this is less likely for the other paths shown below.

As highlighted in red in both Figure 20 and Figure 21, there is a very high probability that both AS3257 (Global Telecom & Technology) and AS2516 (KDDI CORPORATION) can monitor both ends of the Tor communication, thus being able to conduct a correlation attack to deanonymize the client and reveal that the client is using Tor to access Kyodo News. This situation becomes even worse if many ASes collude with one another to compromise the user's anonymity. The worst case scenario is when a government passes laws requesting the collaboration of ISPs to monitor Internet traffic nation-wide.

### 4.3.2 Extra hub + Tor as a potential solution

In order to help Tor users to overcome the "loop phenomenon", one possible solution is to use VPN servers that are located outside of users' home country. Using foreign VPN servers helps to prevent an ISP from knowing whether or not Tor is being used to access a local website, since all of the user's connections are encrypted in the VPN tunnel and the real IP address communicating with the guard node is outside of the user's home country. We suggest the use of VPN Gate [72] because it is a network of volunteer-based public VPN servers that are widely distributed around the world, making it more feasible to implement.

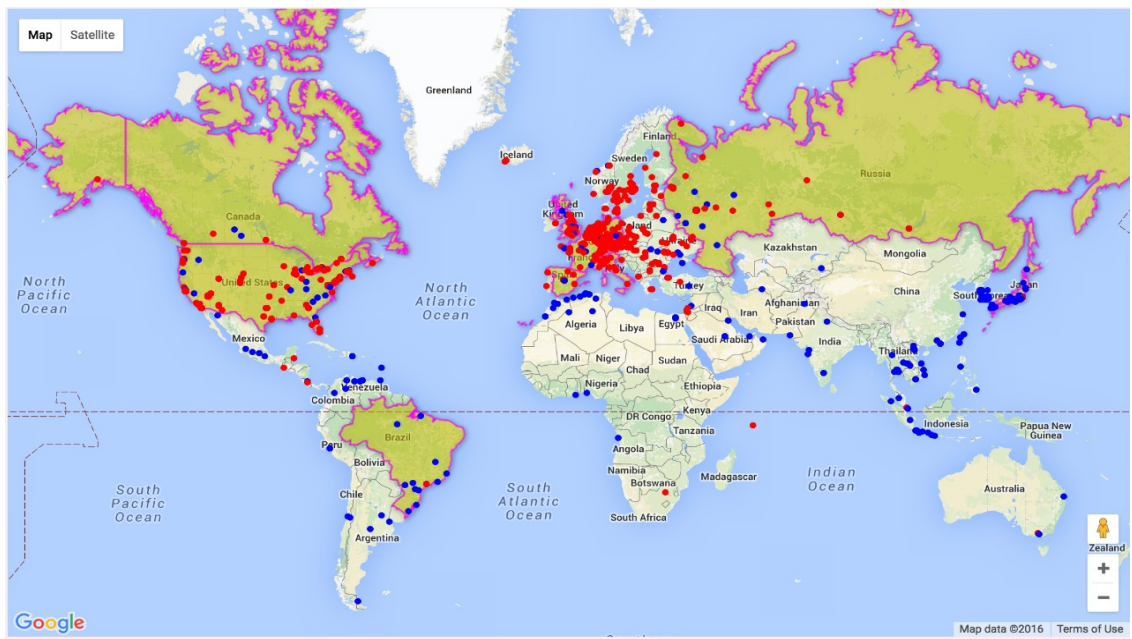

Figure 22: The distribution of VPN Gate servers (blue) and guard nodes (red) over the top 10 countries of daily direct connecting Tor users (highlighted)

# Chapter 5 ASmoniTor

In this chapter, we discuss the implementation of ASmoniTor, which is an autonomous system monitor for mitigating correlation attacks in the Tor network. ASmoniTor is an online platform created to assist Tor users by mitigating the threat of correlation attacks posed by AS-level adversaries while surfing the Internet with Tor Browser Bundle. For clarity, ASmoniTor is ideal for interactive web surfing users, but not for dark web and BitTorrent users.

## 5.1 Design goal of ASmoniTor

First of all, it is necessary to clarify that our ASmoniTor system does not aim to serve all types of Tor users because we do not truly know how many types of users exist within the Tor network. While some use Tor to access websites that are censored in their home country, others use it to protect their privacy and to maintain online anonymity in the midst of state-sponsored surveillance activities. According to [73], as a low-latency anonymous communication tool, Tor is often used for interactive web surfing. Therefore, we designed ASmoniTor to primarily serve this segment of users.

With a simple user interface, ASmoniTor allows an end user to input two parameters: the set of ASes between herself and the guard, and the intended destination host. Based on the anti-Raptor algorithm [60], ASmoniTor will check all of Tor's available Tor exit nodes to determine which nodes are unsafe for the user. Finally, a list of unsafe exit nodes is returned for the user to append to torrc[1], which is a configuration file in the Tor browser. With these unsafe exit nodes appended to the configuration file, the Tor browser will avoid choosing them when creating circuits. As a result, the user is safe from correlation attacks posed by AS-level adversaries.

## 5.2 Lessons learned from previous works

As briefly mentioned in Section 2.9, the ultimate goal of this study is to improve remaining drawbacks of previous works that made their proposed AS-aware Tor

---

[1] Although the location of the torrc file may change in future Tor Browser versions, at the time of composing this study, it is located at \Tor Browser\Browser\TorBrowser\Data\Tor\torrc.

clients less practical and thus not widely used by Tor users.

First, most previous works only published the source code of an AS-aware Tor client, instead of a readily executable program. Substantial effort and specialized knowledge are required to properly compile the source code, which makes these AS-aware clients less accessible to the many non-technical Tor users. Furthermore, the Tor browser is updated almost monthly so the maintenance of these AS-aware Tor clients is cumbersome for the user.

Second, most previously proposed AS-aware clients consist of a snapshot of Internet topology and/or pre-computed AS-paths. Edman *et al.* in [49] proposed a 2 MB packet, containing an AS topology snapshot and a prefix table for running the AS-awareness path selection. LASTor [50] requires Tor users to initially download 13 MB of AS-related data, including inter-AS links, AS three-tuples, and AS path lengths. Astoria [54] was introduced as a complete AS-aware Tor client with a packet size of 88.3 MB when compressed, and 462 MB when extracted. However, due to the dynamic nature of the Internet, the snapshot of the Internet topology has to be continually updated in order to assure near real-time AS path inferences. Thus, it is too complicated to use the above proposals offline.

Third, not all Tor users have the same concern about the threat of AS-level adversaries. In particular, those who only use Tor to bypass censorships of their local ISPs or authorities. Thus, AS-aware path selection should not be implemented in the entire Tor network as proposed in anti-Raptor [60].

Considering these drawbacks, we aim to design ASmoniTor so that it operates independently from the Tor network and is easy for both technical and non-technical users to use. In addition, the novel design of ASmoniTor will help to reduce errors of AS-path inference techniques, improving the accuracy of the AS-aware path selection process.

In most related works, the main issue was the privacy of Tor users while using third party APIs, such as online AS-path inference systems and GeoIP mapping tools, because such platforms may leak information of Tor users. However, in Subsection 5.3, we will demonstrate that ASmoniTor is designed with a high priority on user privacy. If the user follows our instructions, we cannot obtain any information about the user, despite being the developers of ASmoniTor.

## 5.3 System architecture

ASmoniTor acts as a proxy to assist Tor users concerned about correlation attacks to communicate with other third-party APIs, including AS-path inference systems, and GeoIP mapping tools. By collecting data from those APIs, ASmoniTor aims to maintain a database, as shown in Figure 23.

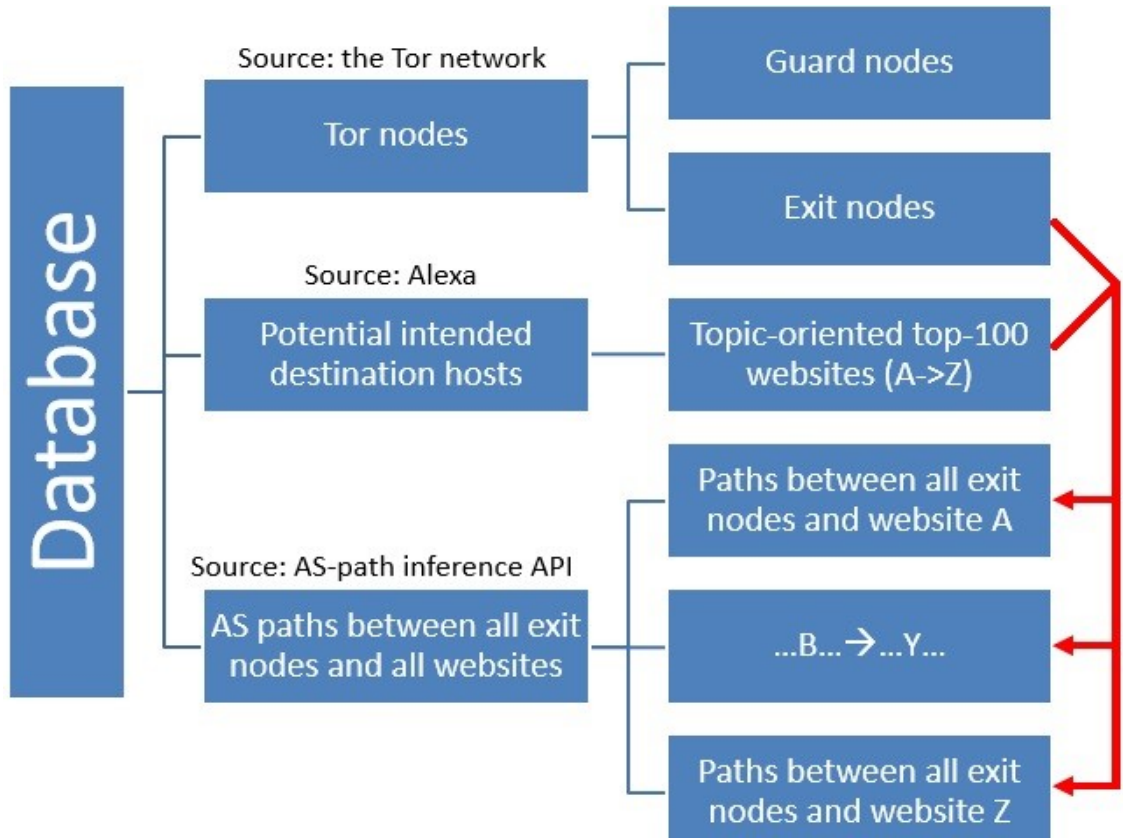

Figure 23: Database of ASmoniTor

As shown in Figure 23, the database of ASmoniTor contains three primary data elements: current Tor nodes, most visited websites, and AS paths between all current exit nodes and all the most visited websites. Since the Tor network publishes a new consensus file every hour, ASmoniTor is designed to fetch the consensus file and update its database when there is a new entry added. Similarly, data of topic-specific websites are fetched and frequently updated from Alexa, as described in subsection 4.2.3.

After obtaining the newest data of both exit nodes and lists of likely visited websites, all AS paths between exit nodes and websites are computed using an AS-path inference API [57]. This dataset comprises of near real-time AS paths because the API uses the daily updated data of Route Views Project [74]. These AS paths are then stored in the database so when the user executes ASmoniTor, the system just needs to intersect the set of ASes inputted by the user with the precomputed AS paths determined by the anti-Raptor algorithm [60].

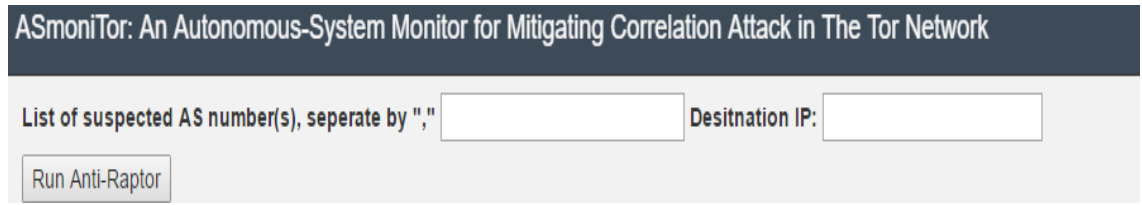

Figure 24: User interface of ASmoniTor

As mentioned in Section 5.2, most previous works were deeply concerned about information leakage to third party APIs if the AS-aware path selection was done online. Therefore, we designed ASmoniTor so that the user can decide to what extent she wants to give away her information. The user just needs to input a list of AS numbers that appear between her PC and the guard node. The user can obtain this list by executing a traceroute measurement on her own machine. The measurement will return a list of ASes similar to Figure 20. In order to protect her information from being traced back or leaked, the user can ***mix up the order of AS numbers*** that she inputs into the first textbox on the ASmoniTor user interface. By mixing up the order, in spite of being the developers of ASmoniTor, we have no way of knowing which guard node the user is using or her real IP address since ASmoniTor can only be accessed using Tor browser.

Although the order of AS numbers is mixed up to increase user anonymity while using ASmoniTor, it does not have any negative impact on the accuracy of the AS-aware path selection process because the anti-Raptor algorithm is not affected by the order of ASes in an AS-path.

One of the key technical contributions of ASmoniTor design is that we let Tor users run traceroute measurements themselves and input the set of ASes obtained directly from their traceroute measurements. This approach is far more reliable than AS-path measurements done by inference techniques because the set of ASes is obtained directly from a real-time control-plane measurement. This feature distinguishes ASmoniTor from prior proposals in which AS-path measurements are solely based on inference techniques (aka data-plane measurements).

It is also necessary to mention that the initial traceroute measurement does not compromise the user's anonymity because all of the entities that appear on the route between the user and her guard know her IP address once she starts using Tor.

Furthermore, since the user in this case is clearly aware of which ASes appear between her and the guard node, she can ***omit one or more ASes that she trusts***. In other words, the user can just input a set of ASes that she suspects may be able to conduct correlation attacks on her Tor communication, instead of all ASes appearing in the traceroute measurement.

To further protect her anonymity, the user can also ***divide the set of ASes into more than one set and query them using ASmoniTor several times***. The user can append a compilation of unsafe exit nodes to her torrc file. This task can be accomplished because Tor browser allows users to change their identity freely (i.e., change the exit node), which prevents ASmoniTor from viewing the complete set of ASes and preserves user anonymity.

For instance, the user in Figure 25 does not trust AS1103 so when she wants to access a website that has an IP address of 141.0.174.41, she inputs these two parameters. ASmoniTor queries the back-end database and runs anti-Raptor algorithm to return a list of exit nodes that should be avoided to prevent AS1103 from carrying out a correlation attack on her traffic.

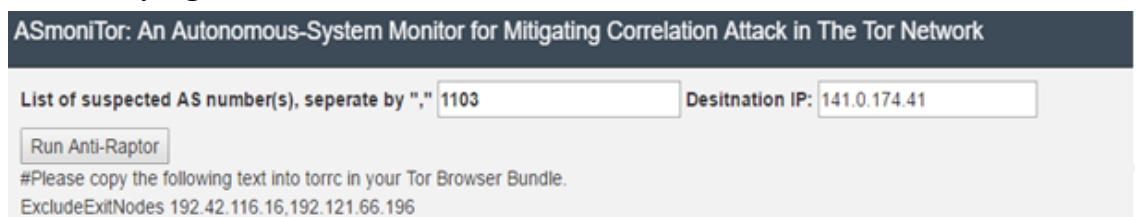

Figure 25: An example of ASmoniTor usage

In more detail, after the user inputs these two parameters, ASmoniTor sends a query to the back-end database to search for a table containing all of the AS paths between all exit nodes and 141.0.174.41 (warning: this is an IP address of an adult website obtained from the top sites in the adult category of Alexa, so please do not try to access it directly from you PC to prevent any claim that may come to you from your ISP or your organization). Next, ASmoniTor searches all of the rows containing AS1103 and notes the IP addresses of exit nodes in those rows. Once the process is finished, ASmoniTor prints the IP addresses of all unsafe exit nodes it identified. In this case, the unsafe exit nodes identified were 192.42.116.16 and 192.121.66.196.[1]

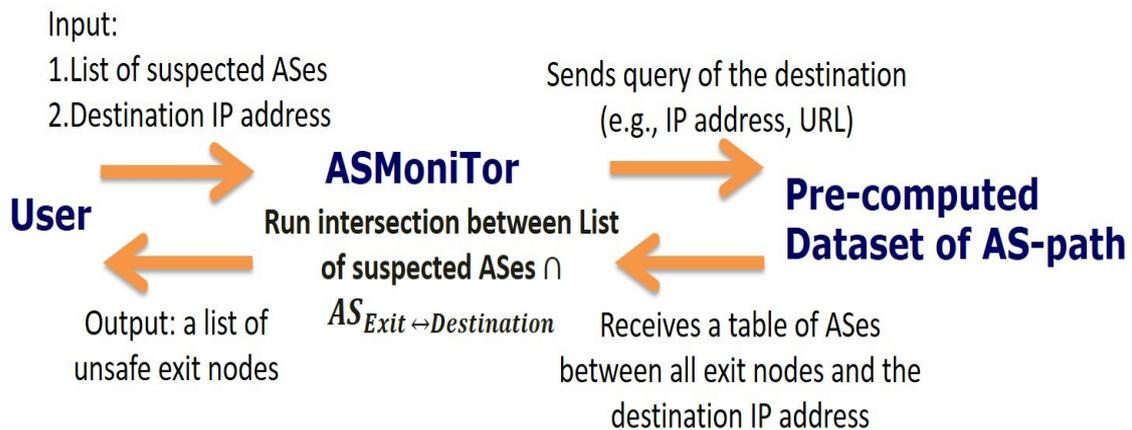

Figure 26: Operating mechanism of ASmoniTor

## 5.4 Limitations

Despite its advantages and improvements mentioned above, ASmoniTor has four main drawbacks that should be addressed in future work.

First, ASmoniTor is still in its trial period and may not be able to handle a large number of simultaneous queries. If too many users try to execute our system at the same time, it may crash unexpectedly. If the system is rated highly by the Tor research community, we will utilize a stronger server and employ SSL (Secure Sockets Layer) to serve a larger number of users.

Second, we initially planned to integrate PathCache into our system so that we could always measure the most up-to-date and near real-time AS paths when deploying the anti-RAPTOR algorithm. PathCache is an integrated system of both data-plane and control-plane measurements, which is superior to previously known AS path inference APIs. However, while rolling out the ASmoniTor system, the PathCache server went down unexpectedly, which forced us to switch back to the old AS path inference technique [57].

---

[1] Note that this result is obtained at the time we fetch the Tor consensus file at 16:30 UCT on July 9th 2016.

Third, although ASmoniTor was able to improve the accuracy of AS-path measurements between Tor users and guard nodes by utilizing real-time traceroute measurements carried out by the user, data-plane measurements still need to be used to predict AS paths between exit nodes and the most likely visited websites. However, despite its flaws, ASmoniTor is superior to previous proposals, which relied solely on data-plane measurements to predict all AS paths. In the next chapter, we will introduce a preliminary investigation that shows how AS-path measurements can be further improved in future work.

Finally, we only focused on developing an AS monitor system to assist Tor users in mitigating the threat of AS-level adversaries. However, due to the nature of the Internet (e.g., interconnection, interdependence, etc.), Tor is still susceptible to two other higher level adversaries: ISP-level adversaries and state-level adversaries. In the case of ISP-level adversaries, an ISP may be able to deanonymize Tor users by making use of all ASes within its control in order to conduct correlation attacks. In the case of state-level adversaries, a government can pass laws requiring ISPs to collectively monitor all Internet traffic passing through the country's network backbone to compromise Tor users. Consequently, ASmoniTor, in its current state, cannot help users to mitigate correlation attacks from such high-level adversaries. However, with future improvements, it may be possible to create a safe Tor path in such circumstances. We plan to improve ASmoniTor by allowing Tor users to input the name of an IPS (e.g., NTT) or country (e.g., Japan) and providing them with a list of unsafe exit nodes whose AS paths pass through the specified ISP or nation.

# Chapter 6 Conclusion and Future Work

## 6.1 Conclusion

We hope that this study will help to raise awareness of some of the issues surrounding the maintenance of Internet user privacy and anonymity. In addition, although Tor is one of the most popular anonymous communication tools, it cannot protect the user's privacy in all cases. We used a real-world example in which Tor is not effective in protecting against AS-level adversaries.

Therefore, in order to help Tor users to become completely anonymous, we extensively studied correlation attacks posed by AS-adversaries, which can be used to deanonymize the Tor user. We analyzed works published over the last decade to determine why none of these previous proposals were successfully taken up by end users. Based on the drawbacks of previous works, we proposed ASmoniTor, an autonomous system monitor, to help Tor users mitigate the threat of correlation attacks posed by AS-level adversaries. While designing ASmoniTor, we were also careful not to cause additional scalability problems in the Tor network.

Although ASmoniTor is still in the early stages of development, we hope that we will be able to distribute it widely to end users once the limitations mentioned in Section 5.4 are resolved.

## 6.2 Future work

For future expansion of ASmoniTor, we would like to take the following issues into account.

### 6.2.1 Integrate VPN into ASmoniTor

As stated in Section 4.3, VPN Gate is a potential solution for the Tor dilemma and can also be utilized when Tor users face the threat of ISP-level or state-level adversaries. We plan to integrate VPN Gate [72] into ASmoniTor so that the program will suggest using public VPN servers when it cannot find many safe exit nodes. After initializing the VPN connection, users can revisit ASmoniTor with their new identity (i.e., IP address of VPN) to check which exit nodes are safe.

### 6.2.2 Cloudfare and Google Captcha problem

Due to the increasing number of negative Tor users in recent years, many websites have employed Cloudfare or Google Captcha extensively. However, such an approach may compromise user privacy and anonymity because it makes the visited destination host become two separate hosts (i.e., the intended host and the host of Cloudfare or Google Captcha server). Therefore, in future work, we plan to test all websites in our database to see which ones are employing Cloudfare and Captcha. We will then measure the AS paths between all exit nodes and Cloudfare/ Captcha servers.

### 6.2.3 A "less-bad" method to measure accurate AS paths

To the best of our knowledge, the most accurate way to obtain the set of ASes appearing between the two ends of Tor anonymous paths is to utilize traceroute-like tools. Nonetheless, due to the widely distributed and volunteer-based nature of the Tor network, it is quite challenging to gain physical access to all guard and exit nodes in order to execute traceroute commands. This was a drawback of most previous works in which the authors used only a limited number of Tor exit nodes, whose owners were willing to participate in their experiments, to carry out the traceroute measurements. For example, Murdoch and Zieliński carried out traceroute experiments with only Tor nodes based in the UK to sample Tor traffic [75]. In a similar manner, Joshua Juen *et al.* [53] requested Tor exit runners to participate in their traceroute experiments, but were only able to obtain traceroute results from 28 exit nodes.

Although we do not have a perfect solution to the current problem of accurately measuring AS paths, based on a preliminary test done to count how many public servers that provide free traceroute services from their servers across the Internet, we expect that we will be able to propose a "less-bad" method of accurately obtaining AS paths between the two ends of Tor anonymous circuits. By analyzing the Tor consensus document, obtained at 16:30 UCT on July 9th, 2016, we found that there are 72 public traceroute servers that host guard nodes and 53 public traceroute servers that host exit nodes. These results are much better than any prior works mentioned above.

# Acknowledgments

First and foremost, I would like to thank Professor Yoshikawa Masatoshi, head of Distributed Information Systems Laboratory for giving me a life-changing chance to study in his lab during the last two years. I would not be who I am today without his admission decision for me to study in one of the most prestigious universities in Japan, Kyoto University.

I also would like to express my sincere gratitude to Professor Asano Yasuhito for his dedicated support, constructive criticism, and insight during the thesis process as well as during my master's study. Also thanks to the academic advice of Professor Tanaka Katsumi and Professor Shinkuma Ryoichi, my last two years at Kyoto University were really fruitful in term of research activities with two papers accepted by international conferences and one journal paper published in the IEEE digital library. Moreover, one of the papers fortunately received the outstanding paper award in the 18th IEEE International Conference on Advanced Communications Technology, and was widely mentioned by medias.

Next, I would like to gratefully acknowledge the financial support of Ministry of Education, Culture, Sports, Science and Technology (MEXT) of Japan, and Japan Society for the Promotion of Science (JSPS) KAKENHI Grant Number 15K00423 and the Kayamori Foundation of Informational Science Advancement during my master's study.

Special thanks to my junior, Mr. Le Quang Thai, a master's candidate at Department of Computer Science of Illinois Institute of Technology, for his technical supports in building up ASmoniTor which is the heart of this study.

Last but not least, I am greatly thankful to Mr. Nathan Kellock. He was very generous to spend a lot of time on helping me correct spelling and grammar mistakes in this thesis. Furthermore, he always encourages and gives me a lot of motivation to move forward whenever I face difficulty.

Thanks all for all the amazing and valuable supports,

Kyoto, August 1st, 2016.

HOANG, Nguyen Phong

# References

[1]    D. Kahn, *The Codebreakers: The Comprehensive History of Secret Communication from Ancient Times to the Internet*. Simon and Schuster, 1996.

[2]    N. P. HOANG, Y. ASANO, and M. YOSHIKAWA, "Your Neighbors Are My Spies: Location and other Privacy Concerns in GLBT-focused Location-based Dating Applications," *ICACT Trans. Adv. Commun. Technol.*, vol. 5, no. 3, pp. 851–860, May 2016.

[3]    B. Rainey, E. Wicks, and C. Ovey, *The European Convention on Human Rights*. Oxford University Press, 2014.

[4]    United Nations, "The Universal Declaration of Human Rights." [Online]. Available:          http://www.un.org/en/universal-declaration-human-rights/index.html.

[5]    N. P. Hoang and D. Pishva, "Anonymous communication and its importance in social networking," in *The 16th IEEE International Conference on Advanced Communication Technology, ICACT*, 2014, pp. 34–39.

[6]    G. Van Blarkom, J. Borking, and J. Olk, "Handbook of privacy and privacy-enhancing technologies," *Priv. Inc. Softw. …*, pp. 42 – 50, 2003.

[7]    S. Harris and J. Hudson, "Not Even the NSA Can Crack the State Dept's Favorite Anonymous Network," *Foreign Policy*, 04-Oct-2013.

[8]    B. Y. J. Scahill and J. Begley, "The Great SIM Heist: How Spies Stole the Keys to the Encryption Castle," *Intercept*, pp. 1–11, Feb. 2015.

[9]    J. Ball, "NSA stores metadata of millions of web users for up to a year, secret files show," *Guard.*, Sep. 2013.

[10]   S. Murai, "NSA whistleblower Snowden says U.S. government carrying out mass surveillance in Japan," *The Japan Times*, 04-Jun-2016.

[11]   E. MacAskill, J. Borger, N. Hopkins, N. Davies, and J. Ball, "GCHQ taps fibre-optic cables for secret access to world's communications," *Theguardian.Com*, Jun. 2013.

[12]   N. P. Hoang and D. Pishva, "A TOR-based anonymous communication

approach to secure smart home appliances," *ICACT Trans. Adv. Commun. Technol.*, vol. 3, no. 5, pp. 517–525, 2015.

[13] "Tor Metrics. Relays and bridges in the network." [Online]. Available: https://metrics.torproject.org/networksize.html.

[14] "Tor Browser." [Online]. Available: https://www.torproject.org/projects/torbrowser.html.en.

[15] "Tails: The amnesic incognito live system." [Online]. Available: https://tails.boum.org/index.en.html.

[16] D. Goldschlag, M. Reed, and P. Syverson, "Hiding Routing information," *Inf. Hiding*, vol. 1174, pp. 137–150, 1996.

[17] M. G. Reed, P. F. Syverson, and D. M. Goldschlag, "Anonymous connections and onion routing," *IEEE J. Sel. Areas Commun.*, vol. 16, no. 4, pp. 482–493, 1998.

[18] D. Goldschlag, M. Reed, and P. Syverson, "Onion Routing," *Commun. ACM*, vol. 42, no. 2, pp. 39–41, 1999.

[19] R. Dingledine, N. Mathewson, and P. Syverson, "Tor: The second-generation onion router," *SSYM'04 Proc. 13th Conf. USENIX Secur. Symp.*, vol. 13, p. 21, 2004.

[20] "Tor Bridges." [Online]. Available: https://www.torproject.org/docs/bridges.html.

[21] R. N. M. Dingledine, "Tor Path Specification," pp. 1–12, 2011.

[22] Tor Project, "Tor directory protocol, version 3." [Online]. Available: https://gitweb.torproject.org/torspec.git/tree/dir-spec.txt.

[23] J. McLachlan, A. Tran, N. Hopper, and Y. Kim, "Scalable onion routing with torsk," *Proc. 16th ACM Conf. Comput. Commun. Secur. CCS 09*, p. 590, 2009.

[24] "Tor Metrics. Direct users by country." [Online]. Available: https://metrics.torproject.org/userstats-relay-country.html.

[25] Lunar, "Tor Weekly News — September 4th, 2013," 2013.

[26] J. R. Douceur, "The Sybil Attack," in *Peer-to-peer Systems*, P. D. Druschel@cs.rice.edu and A. R. Antr@microsoft.com, Eds. Springer Berlin Heidelberg, 2002, pp. 1–6.

[27]    "Tor Weekly News — December 31st, 2014." [Online]. Available: https://blog.torproject.org/blog/tor-weekly-news-%E2%80%94-december-31st-2014.

[28]    "Tor Metrics. Number of bytes spent on answering directory requests." [Online]. Available: https://metrics.torproject.org/dirbytes.html.

[29]    "Collector: data-collecting service in the Tor network." [Online]. Available: https://collector.torproject.org/.

[30]    M. Madden, L. Rainie, K. Zickuhr, M. Duggan, and A. Smith, "Public Perceptions of Privacy and Security in the Post-Snowden Era," *Pew Res. Cent.*, 2014.

[31]    "The Tor Challenge." [Online]. Available: https://www.eff.org/torchallenge/.

[32]    "Noisebridge Tor." [Online]. Available: http://tor.noisebridge.net/.

[33]    "Tor FAQ - How do I decide if I should run a relay?" [Online]. Available: https://www.torproject.org/docs/faq.html.en#HowDoIDecide.

[34]    S. J. Murdoch and G. Danezis, "Low-cost traffic analysis of Tor," in *Proceedings - IEEE Symposium on Security and Privacy*, 2005, pp. 183–195.

[35]    N. S. Evans, R. Dingledine, and C. Grothoff, "A Practical Congestion Attack on Tor Using Long Paths," *18th USENIX Secur. Symp.*, vol. 19, pp. 33–50, 2009.

[36]    L. Øverlier and P. Syverson, "Locating hidden servers," *Proc. - IEEE Symp. Secur. Priv.*, vol. 2006, pp. 100–114, 2006.

[37]    K. Bauer, D. McCoy, D. Grunwald, T. Kohno, and D. Sicker, "Low-Resource Routing Attacks Against Tor," *Work. Priv. Electron. Soc.*, pp. 11 – 20, 2007.

[38]    "PlanetLab." [Online]. Available: http://planet-lab.org.

[39]    D. Herrmann, R. Wendolsky, and H. Federrath, "Website fingerprinting: attacking popular privacy enhancing technologies with the multinomial naïve-bayes classifier," *... 2009 ACM Work. ...*, pp. 31–41, 2009.

[40]    Y. Shi and K. Matsuura, "Fingerprinting attack on the Tor anonymity system," *Lect. Notes Comput. Sci. (including Subser. Lect. Notes Artif. Intell. Lect. Notes Bioinformatics)*, vol. 5927 LNCS, pp. 425–438, 2009.

[41]  A. Panchenko and L. Niessen, "Website fingerprinting in onion routing based anonymization networks," *Proc. 10th …*, p. 103, 2011.

[42]  K. P. Dyer, S. E. Coull, T. Ristenpart, and T. Shrimpton, "Peek-a-boo, I still see you: Why efficient traffic analysis countermeasures fail," in *Proceedings - IEEE Symposium on Security and Privacy*, 2012, pp. 332–346.

[43]  T. Wang and I. Goldberg, "Improved website fingerprinting on Tor," *Proc. 12th ACM Work. Work. Priv. Electron. Soc. - WPES '13*, pp. 201–212, 2013.

[44]  X. Cai, R. Nithyanand, T. Wang, R. Johnson, and I. Goldberg, "A Systematic Approach to Developing and Evaluating Website Fingerprinting Defenses," *Proc. 2014 ACM SIGSAC Conf. Comput. Commun. Secur. - CCS '14*, pp. 227–238, 2014.

[45]  R. Nithyanand, X. Cai, and R. Johnson, "Glove: A Bespoke Website Fingerprinting Defense," in *Proceedings of the 12th Workshop on Privacy in the Electronic Society (WPES)*, 2014.

[46]  T. Wang, X. Cai, R. Nithyanand, R. Johnson, and I. Goldberg, "Effective Attacks and Provable Defenses for Website Fingerprinting," *23rd USENIX Secur. Symp. (USENIX Secur. 14)*, pp. 143–157, 2014.

[47]  X. Cai, R. Nithyanand, T. Wang, R. Johnson, and I. Goldberg, "A Systematic Approach to Developing and Evaluating Website Fingerprinting Defenses," *Proc. 2014 ACM SIGSAC Conf. Comput. Commun. Secur. - CCS '14*, pp. 227–238, 2014.

[48]  N. Feamster and R. Dingledine, "Location Diversity in Anonymity Networks," *ACM Work. Priv. Electron. Soc.*, no. October, pp. 66 – 76, 2004.

[49]  M. Edman and P. Syverson, "AS-awareness in Tor Path Selection," *Comput. Commun. Secur.*, pp. 380 – 389, 2009.

[50]  M. Akhoondi, C. Yu, and H. V. Madhyastha, "LASTor: A Low-Latency AS-Aware Tor Client," *2012 IEEE Symp. Secur. Priv.*, pp. 476–490, 2012.

[51]  A. Johnson, R. Jansen, M. Sherr, and P. Syverson, "Users Get Routed : Traffic Correlation on Tor by Realistic Adversaries," *Proc. 2013 ACM SIGSAC Conf. Comput. Commun. Secur.*, pp. 337–348, 2013.

[52]  Y. Sun, A. Edmundson, L. Vanbever, E. T. H. Zürich, O. Li, J. Rexford, M. Chiang, P. Mittal, A. Edmundson, L. Vanbever, and J. Rexford, "RAPTOR :

Routing Attacks on Privacy in Tor," *USENIX Secur.*, 2015.

[53] J. Juen, A. Johnson, A. Das, N. Borisov, and M. Caesar, "Defending Tor from Network Adversaries: A Case Study of Network Path Prediction," *Proc. Priv. Enhancing Technol.*, vol. 2015, no. 2, pp. 171–187, 2015.

[54] R. Nithyanand, O. Starov, A. Zair, P. Gill, and M. Schapira, "Measuring and mitigating AS-level adversaries against Tor," *Ndss*, pp. 1–12, 2016.

[55] R. Nithyanand, R. Singh, S. Cho, and P. Gill, "Holding all the ASes: Identifying and Circumventing the Pitfalls of AS-aware Tor Client Design," May 2016.

[56] D. L. Chaum, "Untraceable electronic mail, return addresses, and digital pseudonyms," *Commun. ACM*, vol. 24, no. 2, pp. 84–90, Feb. 1981.

[57] J. Qiu and L. Gao, "AS path inference by exploiting known AS paths," in *GLOBECOM - IEEE Global Telecommunications Conference*, 2006.

[58] K. Loesing, S. J. Murdoch, and R. Dingledine, "A case study on measuring statistical data in the Tor anonymity network," *Lect. Notes Comput. Sci. (including Subser. Lect. Notes Artif. Intell. Lect. Notes Bioinformatics)*, vol. 6054 LNCS, pp. 203–215, 2010.

[59] R. Singh and P. Gill, "PathCache: A Path Prediction Toolkit."

[60] N. P. Hoang, Y. Asano, and M. Yoshikawa, "Anti-RAPTOR: Anti routing attack on privacy for a securer and scalable Tor," in *The 17th IEEE International Conference on Advanced Communication Technology, ICACT*, 2015, pp. 147–154.

[61] "Stem: a Python controller library for Tor." .

[62] "Compass." [Online]. Available: https://compass.torproject.org/. [Accessed: 27-Mar-2015].

[63] "The CAIDA AS Relationships Dataset." [Online]. Available: http://www.caida.org/data/as-relationships/. [Accessed: 27-Mar-2015].

[64] "Hurricane Electric Internet Service: BGP Toolkit." [Online]. Available: http://bgp.he.net/. [Accessed: 27-Mar-2015].

[65] K. Zetter, "Meet Monstermind, The NSA Bot That Could Wage Cyberwar Autonomously," *Wired Magazine*, p. 13, Aug-2014.

[66] R. Dingledine, "One fast guard for life (or 9 months)," *7th Work. ...*, 2014.

[67] S. Hahn, "Privacy-preserving ways to estimate the number of tor users," pp. 1–16, 2010.

[68] "Quantcast." [Online]. Available: http://www.quantcast.com/top-sites-1.

[69] "Collection of censorship blockpages as collected by various sources." [Online]. Available: https://github.com/citizenlab/blockpages.

[70] "GlobalVoices Blockpage Gallery." [Online]. Available: https://advox.globalvoices.org/past-projects/blockpages/.

[71] S. Kelly, M. Earp, L. Reed, A. Shahbaz, and M. Truong, "Freedom on the Net 2015: Privatizing Censorship, Eroding Privacy," 2015.

[72] D. Nobori and Y. Shinjo, "VPN gate: A volunteer-organized public vpn relay system with blocking resistance for bypassing government censorship firewalls," *Proc. 11th USENIX Symp.*, 2014.

[73] D. K. McCoy Tadayoshi; Sicker, Douglas, "Shining Light in Dark Places: Understanding the Tor Network," *8th Priv. Enhancing Technol. Symp.*, pp. 63–76, 2008.

[74] D. Meyer and others, "University of oregon route views project." 2005.

[75] S. J. Murdoch and P. Zieliński, "Sampled Traffic Analysis by Internet-Exchange-Level Adversaries," *Priv. Enhancing Technol.*, pp. 167 − 183, 2007.

[76] "MAXMind IP geolocation databases." [Online]. Available: http://dev.maxmind.com/geoip/legacy/geolite/. [Accessed: 27-Mar-2015].

# Appendix

Table 12: Top-100 websites (Source: Alexa)

| 1 | Google.com | 26 | Ebay.com | 51 | Xvideos.com | 75 | Kat.cr |
|---|---|---|---|---|---|---|---|
| 2 | Youtube.com | 27 | Reddit.com | 52 | Google.com.mx | 76 | Blogger.com |
| 3 | Facebook.com | 28 | Google.co.uk | 53 | Stackoverflow.com | 77 | Google.pl |
| 4 | Baidu.com | 29 | Google.com.br | 54 | Apple.com | 78 | Nicovideo.jp |
| 5 | Yahoo.com | 30 | Mail.ru | 55 | Aliexpress.com | 79 | Soso.com |
| 6 | Amazon.com | 31 | T.co | 56 | Fc2.com | 80 | Alibaba.com |
| 7 | Wikipedia.org | 32 | Pinterest.com | 57 | Google.co.kr | 81 | Pixnet.net |
| 8 | Google.co.in | 33 | Amazon.co.jp | 58 | Google.ca | 82 | Go.com |
| 9 | Twitter.com | 34 | Gmw.cn | 59 | Github.com | 83 | Google.com.au |
| 10 | Qq.com | 35 | Google.fr | 60 | Imdb.com | 84 | Amazon.co.uk |
| 11 | Live.com | 36 | Netflix.com | 61 | Ok.ru | 85 | Xinhuanet.com |
| 12 | Taobao.com | 37 | Tmall.com | 62 | Google.com.hk | 86 | Dropbox.com |
| 13 | Bing.com | 38 | 360.cn | 63 | Pornhub.com | 87 | Xhamster.com |
| 14 | Google.co.jp | 39 | Google.it | 92 | Googleusercontent.com | 88 | Google.com.tw |
| 15 | Msn.com | 40 | Microsoft.com | 64 | Whatsapp.com | 89 | Outbrain.com |
| 16 | Yahoo.co.jp | 41 | Onclickads.net | 65 | Jd.com | 90 | Cntv.cn |
| 17 | Linkedin.com | 42 | Google.es | 66 | Diply.com | 91 | Cnn.com |
| 18 | Sina.com.cn | 43 | Paypal.com | 67 | Amazon.de | 93 | Booking.com |
| 19 | Instagram.com | 44 | Sohu.com | 68 | Google.com.tr | 94 | Coccoc.com |
| 20 | Weibo.com | 45 | Wordpress.com | 69 | Rakuten.co.jp | 95 | Ask.com |
| 21 | Vk.com | 46 | Tumblr.com | 70 | Craigslist.org | 96 | Youth.cn |
| 22 | Yandex.ru | 47 | Imgur.com | 71 | Office.com | 97 | Bbc.co.uk |
| 23 | Google.de | 48 | Blogspot.com | 72 | Amazon.in | 98 | Popads.net |
| 24 | Google.ru | 49 | Chinadaily.com.cn | 73 | Tianya.cn | 99 | Twitch.tv |
| 25 | Hao123.com | 50 | Naver.com | 74 | Google.co.id | 100 | Microsoftonline.com |

Due to the sensitivity of topic-oriented websites, we do not append them into this thesis. For further reference, please visit them using the following URLs.

| Politics | http://www.alexa.com/topsites/category/Top/Society/Politics |
|---|---|
| Opinion | http://www.alexa.com/topsites/category/Top/News/Analysis_and_Opinion |
| Media | http://www.alexa.com/topsites/category/Top/News/Media_Industry |
| Activism | http://www.alexa.com/topsites/category/Top/Society/Activism |
| Support Group | http://www.alexa.com/topsites/category/Top/Society/Support_Groups |
| GLBT | http://www.alexa.com/topsites/category/Top/Society/Gay,_Lesbian,_and_Bisexual<br>http://www.alexa.com/topsites/category/Top/Society/Transgendered |
| Adult | http://www.alexa.com/topsites/category/Top/Adult |